\documentclass[a4paper,11pt]{article}
\usepackage{jcappub}
\usepackage{lineno}
\usepackage{cleveref}
\usepackage{bm}
\usepackage{mathtools}
\usepackage{tikz}
\usepackage{adjustbox}
\usepackage{lineno}

\usepackage{tabularx}

\usepackage{xspace}
\usepackage[table,xcdraw]{xcolor}
\usepackage{booktabs}
\usepackage{subcaption}
\usepackage{multirow}

\usepackage{graphicx}

\crefname{equation}{}{}

\definecolor{keyword}{HTML}{008000}
\definecolor{emph}{HTML}{0000FF}
\definecolor{string}{HTML}{A52A2A}
\definecolor{comment}{rgb}{0.0, 0.44, 1.0}
\definecolor{back}{HTML}{F8F8F8}
\definecolor{arrow}{HTML}{745334}

\usepackage{listings}

\lstdefinestyle{cppstyle}{
  language=C++, 
  basicstyle=\small\ttfamily,
  keywordstyle=\color{keyword},
  commentstyle=\itshape\color{comment},
  stringstyle=\itshape\color{string},
  numbers=none,
  breaklines=true,
  breakatwhitespace=false,
  frame=single,
  framerule=0.75pt,
  xleftmargin=.15in,
  xrightmargin=.15in,
  postbreak=\mbox{\textcolor{red}{$\hookrightarrow$}\space},
  columns=fullflexible,
  showstringspaces=false,
  keepspaces=true,
}

\newcommand{\code}{\texttt}
\newcommand{\veff}{V_{\text{eff}}}
\newcommand{\phim}{\phi_{\text{m}}}

\usepackage{microtype}
\usepackage[english]{babel}
\newcommand{\origunderscore}{}
\let\origunderscore\_
\renewcommand{\_}{\allowbreak\origunderscore}

\title{\boldmath \textbf{\code{HydroGrav}}: Precise hydrodynamics and gravitational waves for cosmological phase transitions}

\author[a,1]{Flynn Linton,\note{Corresponding author.}}
\author[a]{William Searle,}
\author[a]{Xiao Wang,}
\author[a]{and Csaba Bal\'{a}zs}
\affiliation[a]{School of Physics and Astronomy, Monash University,\\
Melbourne 3800 Victoria, Australia}

\emailAdd{flynn.linton@monash.edu}
\emailAdd{william.searle@monash.edu}
\emailAdd{xiao.wang1@monash.edu}
\emailAdd{csaba.balazs@monash.edu}

\abstract{We present \href{https://github.com/Hydro-Grav/HydroGrav}{\code{HydroGrav}}, a \code{C++} code used to construct self-similar fluid profiles, using the exact equation of state determined directly from the effective potential, for any particle physics model capable of producing a first-order electroweak phase transition. \code{HydroGrav} also supports the bag and $\mu\nu$ (or improved bag) equations of state and includes an implementation of the sound shell model for computing the corresponding gravitational wave spectra. Using this framework, we compare the fluid profiles and gravitational wave spectra for the simplified (bag and $\mu\nu$) and exact equations of state for a $\mathbb{Z}_2$-symmetric extension of the Standard Model. Furthermore, we perform a scan across the parameter space of this model to identify regions where the simplified and exact equations of state differ in peak amplitude and spectral shape. Finally, we estimate the effect of using the exact equation of state on the signal-to-noise ratio across the parameter space, as measured by LISA after a 4-year mission.}

\begin{document}
\maketitle
% \flushbottom

% Introduction
\section{Introduction}
\label{sec:intro}
Gravitational wave astronomy offers a unique opportunity to study the history of the early Universe.  Unlike electromagnetic probes, gravitational waves propagate essentially unimpeded from their source, carrying information about physical processes that occurred long before the formation of the cosmic microwave background \cite{Witten:1984rs, Kamionkowski:1993fg}.  One of the most promising targets for future space-based observatories, such as LISA~\cite{LISA:2017pwj, LISA:2024hlh}, is a stochastic gravitational wave background generated by a cosmological first-order phase transition~\cite{Caprini:2015zlo, LISACosmologyWorkingGroup:2022jok}.  Such transitions occur in a wide range of theories beyond the Standard Model and provide a potential observational link between particle physics at the electroweak scale and gravitational wave measurements \cite{Mazumdar:2018review, Caprini:2019egz, athron_cosmological_2024}.

Realising this potential requires theoretical predictions whose precision matches the capabilities of upcoming experiments.  The dominant contribution to the gravitational wave signal is expected to arise from sound waves generated by expanding bubbles of the broken phase, making the resulting spectrum highly sensitive to the hydrodynamic evolution of the plasma \cite{hindmarsh_gravitational_2014, hindmarsh_numerical_2015, Hindmarsh:2017gnf}.  In turn, the hydrodynamics depends on the equation of state of the underlying particle-physics model.  Establishing the accuracy of the approximations commonly employed in phase-transition studies, and understanding how they affect predicted gravitational wave signals, is therefore an important step towards precision gravitational wave cosmology.

As the Universe cools, a first-order phase transition proceeds through the nucleation of bubbles of the energetically favoured phase within the surrounding metastable plasma \cite{Kamionkowski:1993fg}.  These bubbles subsequently expand, converting vacuum energy into bulk fluid motion and reheating the surrounding medium \cite{steinhardt_relativistic_1982, Ignatius:1994hydro}.  The interaction between the bubble walls and the plasma generates sound waves that propagate throughout the fluid long after the bubbles have collided.  Numerical simulations indicate that these acoustic excitations are the dominant source of the resulting stochastic gravitational wave background for most electroweak-scale transitions \cite{hindmarsh_gravitational_2014, hindmarsh_numerical_2015, Hindmarsh:2017gnf}.  Consequently, predicting the gravitational wave spectrum requires an accurate description of the hydrodynamic evolution of the plasma, including the thermodynamic properties that determine the structure of the fluid profiles around expanding bubbles.

While the bag and $\mu\nu$ equations of state have become standard tools in analytical and semi-analytical studies of cosmological phase transitions, their accuracy has rarely been tested against fully model-dependent hydrodynamic calculations \cite{espinosa_energy_2010, Leitao:2014pda, Giese:2020znk, wang_energy_2021, Tian:2024ysd}.  Most existing analyses therefore assume that the radiation-dominated approximation captures the relevant fluid dynamics with sufficient precision, despite the fact that the true thermodynamic properties of the plasma can differ significantly from these simplified descriptions.  As gravitational wave observatories, such as LISA, approach the sensitivity required to probe electroweak-scale phase transitions, understanding the impact of these approximations has become increasingly important.

\code{HydroGrav} \cite{hydrograv} was developed to address this issue.  Building on the established framework of self-similar relativistic hydrodynamics and the sound shell model (SSM), the code computes fluid profiles and gravitational wave spectra directly from the exact equation of state extracted from a finite-temperature effective potential.  At the same time, it retains support for the bag and $\mu\nu$ approximations, enabling direct comparisons between simplified and exact treatments.  This makes \code{HydroGrav} both a precision tool for model-dependent predictions and a platform for quantifying the regime of validity of the approximations that underpin much of the existing literature.

A more precise hydrodynamic treatment is of limited value if the resulting gravitational wave spectrum is ultimately mapped onto a fitting formula derived from the very approximations one seeks to avoid.  Thus, in addition to providing a more accurate hydrodynamic treatment, \code{HydroGrav} also improves upon the gravitational wave modelling adopted in much of the existing literature.  The majority of current studies estimate the sound-wave contribution using fitting formulae calibrated to a limited set of numerical simulations \cite{Caprini:2019egz, Caprini:2025mfr}.  

Although these fits are computationally inexpensive and useful for broad phenomenological surveys, they compress the underlying fluid dynamics into a small number of effective parameters, typically the phase-transition strength, wall velocity, and efficiency factor.  As a consequence, they are unable to capture changes in the detailed structure of the fluid profiles arising from different equations of state.  This introduces a conceptual inconsistency when the hydrodynamics are computed using increasingly sophisticated model-dependent treatments, while the resulting gravitational wave spectrum is still evaluated using fitting formulae derived under simplifying assumptions that may no longer hold.

\code{HydroGrav} instead employs the sound shell model, which constructs the gravitational wave spectrum directly from the velocity and energy-density perturbation profiles generated by the hydrodynamic solution \cite{hindmarsh_sound_2018, hindmarsh_gravitational_2019, pol_characterization_2024}.  This preserves the connection between the microscopic thermodynamics of the underlying particle-physics model, the macroscopic fluid motion, and the resulting gravitational wave signal.  By using the same fluid profiles to determine both the hydrodynamic evolution and the gravitational wave source, the calculation remains internally consistent and retains information about the shape and structure of the sound shells that is lost in fitting-formula approaches.  This makes the framework particularly well suited for quantifying how deviations from simplified equations of state propagate through to observable gravitational wave spectra.

A number of public tools are available for studying various aspects of finite-temperature phase transitions.  This includes, but is not limited to \code{CosmoTransitions} \cite{Wainwright:2011CosmoTransitions}, \code{Vevacious} \cite{Camargo-Molina:2013qva}, \code{AnyBubble} \cite{Masoumi:2017trx}, \code{Thermal functions} \cite{Fowlie:2018eiu}, \code{BubbleProfiler} \cite{Athron:2019BubbleProfiler}, \code{SimpleBounce} \cite{Sato:2019wpo}, \code{FindBounce} \cite{Guada:2020xnz}, \code{BSMPT} \cite{Basler:2024aaf}, \code{PhaseTracer}~\cite{Athron:2024xrh} and \code{TransitionSolver} \cite{TranSolver}, \code{DRalgo} \cite{Ekstedt:2022bff}, \code{BubbleDet} \cite{Ekstedt:2023sqc}, \code{WallGo} \cite{Ekstedt:2024fyq}, \code{ELENA} \cite{Costa:2025pew}, \code{CosmoGW} \cite{Caprini:2026nnk}, \code{TransitionListener} \cite{Matuszak:2026xsz}, and \code{LeWRON} \cite{wang2026lewron}. Several frameworks also provide gravitational wave forecasts based on the resulting transition parameters.

To generate our results in this work, we use \code{HydroGrav} as a module within \code{PhaseTracer}, although the former can also be used as a stand-alone code.  We calculate the gravitational wave power spectrum generated by a first-order phase transition in a singlet extension of the Standard Model across the entire region of the parameter space where such a transition can occur.  We perform this calculation using the simplified equations of state (bag and $\mu\nu$ model) and the exact equation of state, derived directly from the particle physics model, to identify the degree to which the precise hydrodynamic treatment modifies the gravitational wave spectrum. 

In our calculations, we employ the simplified steady-state fluid equations to construct self-similar fluid profiles rather than performing a full non-linear treatment of the hydrodynamics.  This approach allows a complete hydrodynamic and gravitational wave calculation to be performed in approximately 4-5 seconds (on a modern single core) per parameter point, making large-scale scans of beyond-the-Standard-Model parameter spaces computationally feasible.  We note that \code{HydroGrav} is capable of running in parallel across the gravitational wave calculation, allowing for much faster run-times.

Following a quick start guide in section \ref{sec:QuickStart}, sections \ref{sec:Veff} and \ref{sec:hydrodynamics} define the effective potential alongside the exact and simplified equation of state models.  Section \ref{sec:hydrodynamics} details our numerical fluid profile solver. Finally, sections \ref{sec:GWs} through \ref{sec:Comparison} describe gravitational wave production and systematically compare the resulting fluid profiles and gravitational wave (GW) spectra across the different equations of state.

\section{Quick start}
\label{sec:QuickStart}

\begin{figure}[t!]
    \centering
    \includegraphics[width=0.8\linewidth]{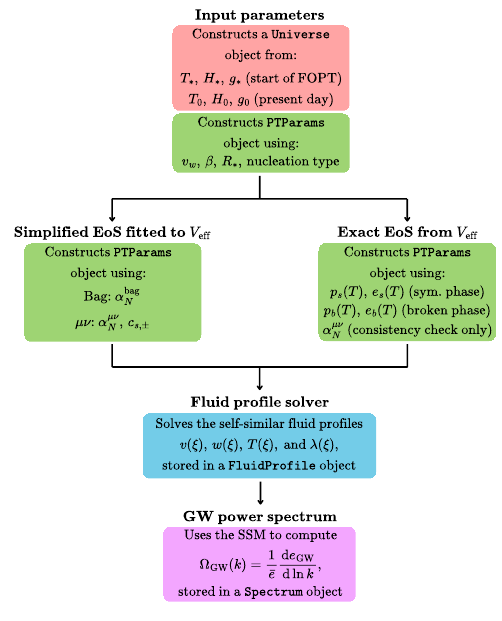}
    \caption{Flow chart describing how the \code{HydroGrav} code constructs the fluid profiles and gravitational wave spectrum, given an underlying particle physics model that produces a first-order electroweak phase transition.}
    \label{fig:flow-chart}
\end{figure}
The results generated in this study were obtained using our new \code{HydroGrav} \code{C++} package \cite{hydrograv}. This code performs precise calculations of the hydrodynamics of phase transitions and calculates the corresponding gravitational wave spectra using the sound shell model, utilising the framework laid out in \crefrange{sec:hydrodynamics}{sec:ssm}. A flow chart of the code's functionality is shown in \Cref{fig:flow-chart}. Starting from a set of phase transition parameters stored in the \code{Universe} and \code{PTParams} classes, and either the bag, $\mu \nu$ (improved bag), or generic $V_{\text{eff}}$ equations of state, the code first solves for the hydrodynamic mode and obtains the self-similar fluid profiles $v(\xi)$, $w(\xi)$, $T(\xi)$, and $\lambda(\xi)$. These are then used to determine the velocity power spectrum, which is ultimately used to obtain the gravitational wave power spectrum. 

Building \code{HydroGrav} requires the following dependencies:
\begin{itemize}
    \item A \code{C++17} compatible compiler,
    \item \href{https://cmake.org/cmake/help/latest/}{\code{CMake}}, version 3.11 or higher,
    \item The \href{https://www.boost.org/}{\code{Boost}} library, version 1.83 or higher,
    \item The \href{https://www.alglib.net/}{\code{ALGLIB}} library, version 4.0 or higher,
    \item The \href{https://www.gnu.org/software/gsl/}{\code{GSL}} library, version 2.7.1 or higher.
\end{itemize}
On Ubuntu or Debian-based distributions, \code{Boost}, \code{ALGLIB} and \code{GSL} can be installed using:
\begin{lstlisting}[language=bash]
$ sudo apt install libboost-all-dev libalglib-dev libgsl-dev  
\end{lstlisting}
On Fedora-based distributions, use the following instead:
\begin{lstlisting}[language=bash]
$ sudo dnf install boost-devel alglib-devel gsl-devel
\end{lstlisting}
Finally, on Mac:
\begin{lstlisting}[language=bash]
$ brew install boost alglib gsl
\end{lstlisting}
Once the required dependencies have been installed, \code{HydroGrav} can be downloaded using:
\begin{lstlisting}[language=bash]
git clone https://github.com/Hydro-Grav/HydroGrav.git
\end{lstlisting}
On a standard UNIX-based system, \code{HydroGrav} can then be built utilising the standard \code{cmake} build system:
\begin{lstlisting}[language=bash]
$ cd HydroGrav 
$ mkdir build
$ cd build 
$ cmake ..
$ make
\end{lstlisting}
This will build \code{HydroGrav} together with all the examples. In addition, we provide the following CMake options, all of which are enabled by default:
\begin{itemize}
    \item \code{ENABLE\_COMPILER\_WARNINGS}. Builds \code{HydroGrav} with compiler warnings enabled.
    \item \code{BUILD\_WITH\_UNIT\_TESTS}. Builds \code{HydroGrav} with unit tests.
    \item \code{BUILD\_WITH\_EXAMPLES}. Builds \code{HydroGrav} with examples.
\end{itemize}
These options can be toggled at configuration time using, for example:
\begin{lstlisting}[language=bash]
$ cmake -D BUILD_WITH_EXAMPLES=ON ..
\end{lstlisting}
The examples included in \code{HydroGrav} are located in the \code{\$example} directory, and if the examples have been built, they can be run using, for example:
\begin{lstlisting}[language=bash]
$ ./bin/run_gw_spectrum
\end{lstlisting}
An example file that describes how to use \code{HydroGrav} to compute fluid profiles and gravitational wave spectra is provided in appendix \ref{app:examples}. Lastly, provided they were built, the unit tests can be run using:
\begin{lstlisting}[language=bash]
$ ./bin/unit_tests
\end{lstlisting}

\section{The effective potential}
\label{sec:Veff}
A cosmological phase transition can occur when the free energy of the system, $\mathcal{F}$, transitions from one phase to a more energetically favoured phase.  The free energy is related to the effective potential \cite{Laine:2016hma}, $\veff (\phi(T), T)$, by
\begin{equation}
    \mathcal{F} = \veff (\phim(T), T) + \mathcal{O} \left ( \frac{ \ln \mathcal{V} }{\mathcal{V}} \right ),
\end{equation}
where $\phim(T)$ is a scalar field configuration that minimises the effective potential at temperature $T$, and the volume term vanishes in the limit of an infinitely large universe ($\mathcal{V} \to \infty$). The effective potential can, in general, be written as the sum of a tree-level and higher-order correction as
\begin{equation}
    \veff(\phi, T) = V_0(\phi) + V_\text{higher order}(\phi, T),
\end{equation}
where temperature dependence enters at the loop level. We remain agnostic about the specific form of $\veff$ and note that the results presented below are applicable to a generic effective potential.
Given a generic effective potential, the thermodynamic pressure, $p(T)$, is obtained from the free energy via 
\begin{equation}
    p(T) = - \mathcal{F}(T) = -\veff (\phim(T), T), \label{eq:thermo-defs-p}
\end{equation}
from which the energy density, $e(T)$, enthalpy, $w(T)$, and entropy, $s(T)$, are derived as
\begin{equation}
    e = T \frac{\partial p}{\partial T} - p, \quad w = e+p = T \frac{\partial p}{\partial T}, \quad s = \frac{\partial p}{\partial T}. \label{eq:thermo-defs}
\end{equation} 
These thermodynamic quantities are essential for the hydrodynamics discussed in section~\ref{sec:hydrodynamics}.

\subsection{Scalar Singlet Extension}
In this work, we demonstrate our improved hydrodynamic calculation using the simplest extension of the Standard Model Higgs sector. Namely, we consider a first-order phase transition driven by a real scalar singlet coupled to the Higgs~\cite{Choi:1993cv, Ham:2004cf}, whose tree-level potential is
\begin{equation}
    V(H, S) = -\mu_h^2 H^\dagger H + \lambda_h (H^\dagger H)^2 - \frac{\mu_s}{2} S^2 + \frac{\lambda_s}{4} S^4 + \frac{\lambda_{hs}}{2} H^\dagger H S^2, \label{eq:Veff-xSM}
\end{equation}
where the Higgs doublet and scalar singlet are
\begin{equation}
    H = \frac{1}{\sqrt{2}} \begin{pmatrix} G^{\pm} \\ \phi_h - h - iG^0 \end{pmatrix}, \quad S = \phi_s + s.
\end{equation}
Here, $\phi_h$ and $\phi_s$ are background fields, $h$ and $s$ are the dynamic degrees of freedom of the Higgs and scalar fields, respectively, and $G^{\pm,0}$ are the Goldstone bosons. 
After symmetry-breaking, we are left with two singly charged  $G^\pm$ and a neutral CP-odd $G^0$ Goldstone bosons, corresponding to the electroweak $W^\pm$ and $Z$ bosons. In addition, we have the CP-even scalars $h$ and $s$, with masses
\begin{equation}
    m_h^2 = 2 v_h^2 \lambda_h, \quad m_s^2 = - \mu_S^2 + \frac{1}{2} v_h^2 \lambda_{hs},
\end{equation}
where $h$ is the SM Higgs, and we have assumed the EWSB vacuum $(\phi_h, \phi_s) = (v_h, 0)$. In regions of the parameter space, the final vacuum is reached by a two-step transition
\begin{equation}
    (\phi_h, \phi_s) \rightarrow (0, v_s) \rightarrow (v_h, 0),
\end{equation}
and we consider GWs generated by the second step of the transition. 

The effective potential is constructed in the context of three-dimensional effective field theory (3dEFT) using the code~\code{DRalgo} \cite{Ekstedt:2022bff}\footnote{\code{DRalgo} utilises the package \code{GroupMath} \cite{Fonseca:2020vke} during model creation.}. 
Specifically, we consider the effective field theory obtained by integrating out the heavy thermal scale, yielding a dimensionally reduced three-dimensional theory. 
Within this framework, the one-loop effective potential is given by
\begin{equation}
    V_{3\text{eff}}(\phi_3, T) = V_{3,0} + V_{3,1} = V_{3,0} - \frac{1}{12 \pi} \sum_{i = h,s} m_i^3(\phi_3),
    \label{eq:3d_veff}
\end{equation}
where the subscript `$3$' indicates quantities in the dimensionally reduced theory. 
The tree-level potential $V_{3,0}$ is obtained by replacing all quantities in $V_0(\phi)$ with their 3D counterparts.
Through matching to the parent UV theory, all 3D parameters become functions of both temperature and the parameters of the underlying 4D Lagrangian. 
The scalar mass eigenstates contributing to the sum in \cref{eq:3d_veff} are obtained using the same approach.

% Relativistic hydrodynamics
\section{Relativistic hydrodynamics}
\label{sec:hydrodynamics}

For a first-order phase transition driven by a scalar field $\phi$, the total stress-energy tensor of the system is 
\begin{equation}
    T^{\mu \nu} = T^{\mu \nu}_{\text{f}} + T^{\mu \nu}_{\phi} ,
\label{eq:FOPT-EM-total}
\end{equation}
where 
\begin{subequations}
\begin{align}
    T^{\mu \nu}_{\mathrm{f}} &= w u^{\mu} u^{\nu} - g^{\mu \nu} p, \label{eq:EM-perfect-fluid} \\
    T^{\mu \nu}_{\phi} &= (\partial^{\mu} \phi) (\partial^{\nu} \phi) - \frac{1}{2} (\partial \phi)^2 g^{\mu\nu}.
\end{align}
\end{subequations}
Here, $T^{\mu \nu}_{\text{f}}$ is the contribution from the relativistic plasma of Standard Model particles, modelled as a perfect fluid in local thermal equilibrium, while $T^{\mu \nu}_{\phi}$ represents the contribution from the scalar field. 
The dynamics of this coupled scalar-fluid system is governed by the conservation of the stress-energy tensor,
\begin{equation}
    \partial_\mu T^{\mu\nu} = \partial_\mu T_{\rm f}^{\mu\nu} + \partial_\mu T_\phi^{\mu\nu} = 0\,.
\end{equation}
In practice, the system can be divided into three distinct regions~\cite{Wang:2023jto}: the bubble wall, the symmetric phase, and the broken phase.
Across the wall, the scalar field dynamics and hydrodynamics are strongly coupled.
However, sufficiently far from the bubble wall, the hydrodynamics dominate, and the field $\phi$ is approximately constant, so $\partial_{\mu} T_{\phi}^{\mu \nu} \approx 0$. Consequently, the conservation of the total stress-energy tensor simplifies to 
\begin{equation}
    \partial_\mu T^{\mu\nu} \approx \partial_{\mu} T^{\mu \nu}_{\text{f}} \approx 0, \label{eq:EM-cons}
\end{equation}
which governs the macroscopic equations of motion for the fluid. 
In typical cosmological phase transitions, the thickness of the fluid shell surrounding the expanding wall is much larger than the bubble wall width.
Therefore, it is sufficient to study the hydrodynamics far from the wall.

After bubble nucleation, the bubble quickly reaches a steady state~\cite{laine_bubble_1994}. Consequently, fluid profiles become self‑similar, preserving their shape while simply rescaling as the bubble expands. 
Assuming a spherically symmetric expansion of bubbles in the fluid, it is then convenient to adopt the radial self-similar coordinate $\xi = r/t$. In this coordinate system, the conservation equation \eqref{eq:EM-cons} reduces to 
\begin{subequations}
\begin{align}
    (\xi - v) \frac{\partial_{\xi} e}{w} &= 2 \frac{v}{\xi} + \left[ 1 - \gamma^2 v (\xi - v) \right] \partial_{\xi} v, \label{eq:eom-1a} \\
    (1 - \xi v) \frac{\partial_{\xi} p}{w} &= \gamma^2 (\xi - v) \partial_{\xi} v. \label{eq:eom-1b}
\end{align}
\end{subequations}
Because the derivatives $\partial_\xi e$ and $\partial_\xi p$ are related to the speed of sound in the plasma 
\begin{equation}
    c_s^2 (T) \equiv \frac{\mathrm{d}p}{\mathrm{d}e} = \left(\frac{\partial p}{\partial \xi}\frac{\partial \xi}{\partial T}\right)\left(\frac{\partial e}{\partial \xi}\frac{\partial \xi}{\partial T}\right)^{-1},
\end{equation}
it is conventional to rewrite these equations of motion in terms of the fluid's velocity, enthalpy, and temperature
\begin{subequations}
\begin{align}
    2 \frac{v}{\xi} &= \gamma^2 (1-\xi v) \left[ \frac{\mu^2(\xi,v)}{c_s^2(T)} - 1 \right] \partial_{\xi} v \label{eq:eom-2a} ,\\
    \frac{\partial_{\xi} w}{w} &= \left[ 1 + \frac{1}{c_s^2(T)} \right] \mu(\xi,v) \gamma^2 \partial_{\xi} v \label{eq:eom-2b} ,\\
    \frac{\partial_{\xi} T}{T} &= \gamma^2 \mu(\xi,v) \partial_{\xi} v \label{eq:eom-2c} .
\end{align}
\label{eq:eom-2}
\end{subequations}
Here $\mu(\xi,v) = \frac{\xi - v}{1 - \xi v}$ represents the Lorentz transformation between the wall frame and the centre-of-bubble frame.
Boundary conditions both ahead of and behind the bubble wall, as well as at any possible shock front, are then required to solve this system.

As the transition proceeds and the bubbles grow to macroscopic scales, the curvature of the bubble wall can be neglected, allowing it to be treated as locally planar. For a wall propagating along the $z$-direction, the fluid flow is purely longitudinal, leaving $T^{00}$, $T^{z0}$, and $T^{zz}$ as the only non-vanishing components of the stress-energy tensor.
Parametrising the fluid four-velocities as $u^{\mu} = \gamma_+ (1,0,0,v_+)$ in the symmetric phase and $u^{\mu} = \gamma_- (1,0,0,v_-)$ in the broken phase, and substituting into \eqref{eq:EM-perfect-fluid}, we obtain the relevant components of the stress-energy tensor,
\begin{equation}
    T_{\text{f}}^{z0} = w_{\pm} v_{\pm} \gamma_{\pm}^2, \quad T_{\text{f}}^{zz} = w_{\pm} v_{\pm}^2 \gamma_{\pm}^2 + p_{\pm}.
\end{equation}
Here, we denote the quantities immediately behind and ahead of the wall with subscripts $-$ and $+$.
The conservation equations of the total stress-energy across the wall therefore yield 
\begin{subequations}
\begin{align}
    \partial_0 T_\phi^{00} - \partial_z T_\phi^{z0} &= -\partial_0 T_{\text{f}}^{00} + \partial_z T_{\text{f}}^{z0} ,\\
    \partial_0 T_\phi^{0z} - \partial_z T_\phi^{zz} &= -\partial_0 T_{\text{f}}^{0z} + \partial_z T_{\text{f}}^{zz} .
\end{align}
\end{subequations}
Neglecting the effects of cosmic expansion over the short duration of the transition, the bubble wall rapidly asymptotes to a steady-state configuration characterised by a constant terminal velocity. Consequently, once this steady state is reached, all explicit time derivatives vanish in the rest frame of the bubble wall, and the conservation equations simplify to 
\begin{equation}
    \partial_z T_{\text{f}}^{z0}  + \partial_z T_\phi^{z0}= \partial_z T_{\text{f}}^{zz} + \partial_z T_\phi^{zz} = 0. \label{eq:matching-conservation}
\end{equation}
Integrating the conservation equation \eqref{eq:matching-conservation} across the bubble wall yields the standard hydrodynamic continuity (or matching) conditions
\begin{equation}
    w_+ v_+ \gamma_+^2 = w_- v_- \gamma_-^2, \quad w_+ v_+^2 \gamma_+^2 + p_+ = w_- v_-^2 \gamma_-^2 + p_-.
\label{eq:matching-conds}
\end{equation}
Equivalently, these can be rewritten as
\begin{equation}
    v_+ v_- = \frac{p_+ - p_-}{e_+ - e_-}, \quad \frac{v_+}{v_-} = \frac{e_- + p_+}{e_+ + p_-}.
\label{eq:matching-conds-2}
\end{equation}
Hence, hydrodynamic solutions for a first-order phase transition must satisfy \eqref{eq:eom-2}, subject to the matching conditions \eqref{eq:matching-conds} (or \eqref{eq:matching-conds-2}) at the phase boundary.

A parallel argument applies to hydrodynamic configurations that develop a shock front at $\xi_{\mathrm{sh}}$. 
Enforcing energy-momentum conservation \eqref{eq:EM-cons} imposes a similar set of matching conditions across the shock front,
\begin{equation}
    w_2 v_2 \gamma_2^2 = w_1 v_1 \gamma_1^2, \quad w_2 v_2^2 \gamma_2^2 + p_2 = w_1 v_1^2 \gamma_1^2 + p_1, \label{eq:matching-conds-sh}
\end{equation}
or equivalently,
\begin{equation}
    v_2 v_1 = \frac{p_2 - p_1}{e_2 - e_1}, \quad \frac{v_2}{v_1} = \frac{e_1 + p_2}{e_2 + p_1}.
\label{eq:matching-conds-sh-2}
\end{equation}
Here, we denote quantities immediately behind and ahead of the shock with subscripts 1 and 2, respectively. Beyond the shock, the plasma remains unperturbed ($\tilde{v}_2=0$), which fixes $v_2=\xi_{\text{sh}}$, $w_2=w_N$, and $T_2=T_N$. The resulting fluid profiles $v(\xi)$, $w(\xi)$ and $T(\xi)$ dictate the shape and amplitude of the corresponding gravitational wave spectrum, as we will see in \cref{sec:GWs}.
To proceed further and obtain these fluid profiles, the final required ingredient is the equation of state.

\subsection{Equation of state}
For a generic theory that undergoes a single first-order phase transition, the effective potential possesses two distinct minima: $\phi_m^+$, associated with the high-temperature symmetric phase, and $\phi_m^-$, corresponding to the low-temperature broken phase. Utilising equations \eqref{eq:thermo-defs-p} and \eqref{eq:thermo-defs}, the equations of state in the symmetric ($s$) and broken ($b$) phases are expressed as 
\begin{equation}
\begin{aligned}
    p_s(T) &= -V_{\text{eff}}(\phi_m^+ (T)), T), \quad &&e_s(T) = -T \frac{\partial V_{\text{eff}}}{\partial T} + V_{\text{eff}}(\phi_m^+ (T)), T) ,\\
    p_b(T) &= -V_{\text{eff}}(\phi_m^- (T)), T), \quad &&e_b(T) = -T \frac{\partial V_{\text{eff}}}{\partial T} + V_{\text{eff}}(\phi_m^- (T)), T) ,
    \label{eq:thermo-defs-veff}
\end{aligned}
\end{equation}
while the associated sound speed in each vacuum phase is given by 
\begin{equation}
    c_{s,s/b}^2(T) = \frac{\mathrm{d}p_{s/b}}{\mathrm{d}e_{s/b}}.
\end{equation}
We denote the pressure and energy density on either side of the wall as $p_{\pm}$ and $e_{\pm}$, defined by
\begin{equation}
\begin{aligned}
    p_+ &= p_s(T_+), \quad e_+ = e_s(T_+),\\
    p_- &= p_b(T_-), \quad e_- = e_b(T_-).
\end{aligned}
\label{eq:EoS-exact}
\end{equation}
For hydrodynamic configurations that develop a shock front, the local thermodynamic quantities at the interface are given by 
\begin{equation}
    \begin{aligned}
        p_2 &= p_s(T_2), \quad e_2 = e_s(T_2) ,\\
        p_1 &= p_s(T_1), \quad e_1 = e_s(T_1) .
    \end{aligned}
    \label{eq:EoS-exact-shock}
\end{equation}
However, a standard approach in the literature approximates the exact equation of state \eqref{eq:thermo-defs-veff}, derived from the effective potential, via the so-called bag model to simplify the hydrodynamics of cosmological phase transitions \cite{steinhardt_relativistic_1982, Kosowsky:1992rz, Kamionkowski:1993fg, Megevand:2012rt, espinosa_energy_2010, Megevand:2009ut, Leitao:2010yw, Leitao:2015fmj, Leitao:2015ola, Konstandin:2010dm}. Within this framework, the thermal contribution to the effective potential is expressed as $(1/3) a(T) T^4$, where $a(T)=g_{\text{eff}}(T) \pi^2/30$ and $g_{\text{eff}}(T)$ denote the effective number of relativistic degrees of freedom at temperature $T$. This assumes a purely radiation-dominated equation of state $p=(1/3)e$. Consequently, the bag equation of state is characterised by a constant sound speed of $c_s^2=1/3$ in both vacuum phases, yielding \cite{chodos_new_1974}
\begin{equation}
\begin{aligned}
    p_+ &= \frac{1}{3} a_+ T_+^4 - \epsilon, \quad &&e_+ = a_+ T_+^4 + \epsilon, \\
    p_- &= \frac{1}{3} a_- T_-^4, \quad &&e_- = a_- T_-^4, \label{eq:bag-eos}
\end{aligned}
\end{equation}
where $\epsilon$ is the false-vacuum energy of the Higgs potential (often referred to as the bag constant), defined to vanish in the broken phase at $T=0$, and $a_{\pm} = g_{\pm} \pi^2 / 30$. The quantities $g_{\pm}$ are the effective degrees of freedom in the symmetric and broken phases, which satisfy $g_+ > g_-$.

The phase transition strength is conventionally parametrised by
\begin{equation}
    \alpha_{\theta} = \frac{\Delta \theta}{3w_+} = \frac{\epsilon}{a_+ T_+^4}, \label{eq:alpha-bag}
\end{equation}
which defines the ratio of the latent vacuum energy released during the transition to the background radiation energy density. Here, the trace anomaly of the fluid's stress-energy tensor is defined as
\begin{equation}
    \theta = g_{\mu\nu} T^{\mu\nu}_{\text{f}} = e-3p,
\end{equation}
with $\Delta \theta = \theta_+ - \theta_-$.

The bag equation of state is generally a robust approximation in the symmetric phase, where Standard Model particles remain effectively massless during an electroweak-scale transition and behave dynamically akin to radiation. However, this approximation typically breaks down in the broken phase, where thermal corrections induce deviations in the sound speed from $c_s^2=1/3$ \cite{Leitao:2014pda, Tenkanen:2022tly}. A phenomenological remedy involves modifying the equation of state \eqref{eq:bag-eos} to accommodate different sound speeds in the respective vacuum phases \cite{Leitao:2014pda, Giese:2020rtr, Giese:2020znk, wang_energy_2021, Ai:2023see}. To this end, we adopt the $\mu\nu$ (or improved bag) model introduced in \cite{wang_energy_2021}\footnote{The formulation of the $\mu\nu$ model described in \cite{Giese:2020znk, Ai:2023see} is physically equivalent to \eqref{eq:improved-bag-eos} when defining the temperature-dependent coefficients in equation (2.15) of \cite{Giese:2020znk} as $a_+(T)=a_+ T^{4-\mu}$ and $a_-(T)=a_- T^{4-\nu}$, with $\mu=1+1/c_{s,+}^2$ and $\nu=1+1/c_{s,-}^2$.}, which reads
\begin{equation}
\begin{aligned}
    p_+ &= c_{s,+}^2 a_+ T_+^4 - \epsilon, \quad &&e_+ = a_+ T_+^4 + \epsilon, \\
    p_- &= c_{s,-}^2 a_- T_-^4, \quad &&e_- = a_- T_-^4, \label{eq:improved-bag-eos}
\end{aligned}
\end{equation}
where $c_{s,+} = c_{s,s}(T_+)$ and $c_{s,-} = c_{s,b}(T_-)$ denote the sound speeds in the symmetric and broken vacuum phases, respectively, and are treated as constants.

In this generalised framework, the trace anomaly takes the form $\bar{\theta}=e-p/c_{s,-}$, yielding the modified strength parameter
\begin{equation}
    \alpha_{\bar{\theta}} = \frac{\Delta \bar{\theta}}{3w_+} = \frac{1}{3(1+c_{s,+}^2)} \left[ 1-\frac{c_{s,+}^2}{c_{s,-}^2} + \left(1+\frac{1}{c_{s,-}^2}\right) \frac{\epsilon}{a_+ T_+^4} \right].
\label{eq:alpha-munu}
\end{equation}
In the limit $c_{s,+}^2 = c_{s,-}^2 = 1/3$, the bag model definition \eqref{eq:alpha-bag} is recovered.

By combining this strength parameter with the matching conditions \eqref{eq:matching-conds-2} and the $\mu\nu$ equation of state \eqref{eq:improved-bag-eos}, the fluid velocity immediately in front of the wall is
\begin{align}
    v_+ = \frac{1}{1+3c_{s,-}^2 \alpha_+} \left[\vphantom{\rule{0pt}{4.9ex}}\right.
    &\left(\frac{v_-}{2} + \frac{c_{s,-}^2}{2v_-} \right) \nonumber \\ &\pm \sqrt{\left(\frac{v_-}{2} + \frac{c_{s,-}^2}{2v_-}\right)^2 + 9c_{s,-}^4 \alpha_+^2 + 3c_{s,-}^2(1-c_{s,-}^2) \alpha_+ - c_{s,-}^2} \left.\vphantom{\rule{0pt}{4.9ex}}\right],
    \label{eq:vp-from-matching}
\end{align}
where $\alpha_+$ is the strength parameter evaluated at $T_+$, and the positive (negative) root is taken for $|v_+|>|v_-|$ ($|v_+| < |v_-|$). For hydrodynamic profiles exhibiting a shock front, the equation of state across the shock \eqref{eq:EoS-exact-shock} takes the form
\begin{equation}
    \begin{aligned}
        p_2 &= c_{s,+}^2 a_+ T_2^4 - \epsilon, \quad &&e_2 = a_+ T_2^4 + \epsilon , \\
        p_1 &= c_{s,+}^2 a_+ T_1^4 - \epsilon, \quad &&e_1 = a_+ T_1^4 + \epsilon , \label{eq:improved-bag-eos-sh}
    \end{aligned}
\end{equation}
so the matching conditions \eqref{eq:matching-conds-sh-2} are
\begin{equation}
    v_2 v_1 = c_{s,+}^2, \quad \frac{v_2}{v_1} = \frac{(T_1/T_2)^4 + c_{s,+}^2}{c_{s,+}^2 (T_1/T_2)^4 + 1}.
\label{eq:shock-matching}
\end{equation}
    
Using either of these equation of state approximations significantly simplifies the relativistic hydrodynamics required to compute the gravitational wave spectrum. This permits a straightforward mapping of a specific effective potential onto the fluid dynamics by extracting the relevant phase transition parameters and evaluating either \eqref{eq:alpha-bag} or \eqref{eq:alpha-munu}. Nevertheless, this methodology remains intrinsically `model-independent', as the parametrised equation of state relies purely on macroscopic properties rather than the microscopic structure of a particular particle physics model.

In the remainder of this work, we quantify the impact of these simplifying assumptions on both the fluid dynamics and the resulting gravitational wave spectrum. We achieve this by performing the hydrodynamic calculations utilising the exact, fully model-dependent equation of state derived directly from the effective potential. While partially model-dependent hydrodynamic treatments have been explored previously \cite{Tian:2024ysd,Wang:2024slx,Wang:2022lyd}\footnote{In \cite{Tian:2024ysd}, the exact equation of state was used within the broken phase, while the $\mu\nu$ equation of state was used in the symmetric phase. In \cite{Wang:2022lyd}, the authors extend the bag model with additional temperature-dependent terms in the broken phase that reflect the structure of the effective potential.}, our fully model-dependent approach operates independently of both the bag and $\mu\nu$ approximations.

    \subsection{Types of hydrodynamic solutions}

    Hydrodynamic solutions for a perfect fluid can be categorised by the direction of flow of the fluid across the bubble wall \cite{landau1987fluid}. For deflagration solutions, the fluid flows outwards from the bubble wall, so the pressure decreases ($p_+>p_-$) and velocity increases ($|v_+| < |v_-|$). For detonation solutions, the fluid flows inwards from the bubble wall, so the pressure increases ($p_+<p_-$) and velocity decreases ($|v_+| > |v_-|$). Here, we use the `$+$' subscript to denote quantities in front of the bubble wall and the `$-$' subscript for quantities behind the wall. The fluid velocities $v_{\pm}$ are given in the bubble wall frame, and we use a tilde to denote the fluid velocities $\tilde{v}_{\pm}$ in the centre-of-bubble frame (also called the fluid frame).    
    
    These solutions can be further categorised as weak (fluid velocities on either side of the wall are both subsonic or both supersonic), strong (one of $v_{\pm}$ is subsonic and the other is supersonic), or Chapman-Jouguet ($v_-=c_{s,-}$ or $v_+=c_{s,+}$). The sixteen types of hydrodynamic solutions are shown in \Cref{tbl:hydrodynamic-sols}. Note that weak supersonic deflagrations and weak subsonic detonations belong to the class of `inverse' phase transitions \cite{barni_hydrodynamics_2024, Barni-inverse-pts}, which we do not consider in this work. These involve the vacuum undergoing a phase transition from a low-temperature phase to a high-temperature phase, so the false vacuum energy is negative (i.e. $\alpha_{\bar{\theta}}<0$) and fluid flows outwards from the bubble in the centre-of-bubble frame. 

    \begin{table}[h!]
        \centering
        \footnotesize% fontsize
        \begin{tabularx}{\textwidth}{|X|X|X|X|X|X|X|}
        \hline
         & Mode & Weak\newline subsonic & Weak\newline supersonic & Strong & Chapman-Jouguet \\ \hline
        \multirow{2}{*}[-0.9\normalbaselineskip]{$v_+ < v_-$} & Deflagration\newline ($\alpha>0$) & \multirow{4}{*}[-2.0\normalbaselineskip]{\shortstack{$v_+ < c_{s,+}$,\\ $v_- < c_{s,-}$}} & \multirow{4}{*}[-2.0\normalbaselineskip]{\shortstack{$v_+ > c_{s,+}$,\\ $v_- > c_{s,-}$}} & \multirow{2}{*}[-1.0\normalbaselineskip]{\shortstack{$v_+ < c_{s,+}$,\\ $v_- > c_{s,-}$}} & $v_+ < c_{s,+}$,\newline $v_- = c_{s,-}$ \\ \cline{2-2} \cline{6-6} 
         & Inv. Deflag.\newline ($\alpha<0$) &  &  &  & $v_+ = c_{s,+}$,\newline $v_- > c_{s,-}$ \\ \cline{1-2} \cline{5-6} 
        \multirow{2}{*}[-0.9\normalbaselineskip]{$v_+ > v_-$} & Detonation\newline ($\alpha>0$) &  &  & \multirow{2}{*}[-1.0\normalbaselineskip]{\shortstack{$v_+ > c_{s,+}$,\\ $v_- < c_{s,-}$}} & $v_+ > c_{s,+}$,\newline $v_- = c_{s,-}$ \\ \cline{2-2} \cline{6-6} 
         & Inv. Deton.\newline ($\alpha<0$) &  &  &  & $v_+ = c_{s,+}$,\newline $v_- < c_{s,-}$ \\ \hline
        \end{tabularx}
        \caption{The sixteen possible hydrodynamic solutions for an expanding bubble in a perfect fluid. For both direct ($\alpha > 0$) and inverse ($\alpha < 0$) cosmological phase transitions, strong deflagrations and detonations are forbidden. Weak supersonic deflagrations and weak subsonic detonations are only possible for inverse phase transitions, whilst weak subsonic deflagrations and weak supersonic detonations are only possible for direct phase transitions. The physical solutions for a cosmological phase transition are shown in \Cref{fig:bag-sol-space}.}
        \label{tbl:hydrodynamic-sols}
    \end{table}
    
    For deflagration solutions, the fluid inside the bubble is at rest in the centre-of-bubble frame ($\tilde{v}_-=0$), so $v_-=\xi_w$. The bubble wall is therefore subsonic ($\xi_w < c_{s,-}$) for weak subsonic deflagrations, or equal to the speed of sound ($\xi_w=c_{s,-}$) for Chapman-Jouguet deflagrations. Strong deflagrations are known to be mechanically unstable \cite{landau1987fluid} and would quickly decay into a rarefaction wave and a weak subsonic deflagration \cite{laine_bubble_1994}. The subsonic expansion of the bubble compresses the fluid in front of the wall, forming a discontinuous shock front $\xi_{\text{sh}} > c_{s,+}$ that reheats the fluid behind it (i.e. $T_+ > T_N$). 
    
    The condition $|v_+|<|v_-|$ for deflagrations gives both a lower and upper bound for the strength parameter $\alpha_+$ such that a shock can form. 
    Inverting equation~\eqref{eq:vp-from-matching}, we find that $\alpha_+^{\text{min}}=0$ for $v_+=v_-$, corresponding to the case in which no energy is transferred from the phase transition to the fluid, and $\alpha_+^{\text{max}}=1/3$ for $v_+=0$, which represents the maximum phase transition strength compatible with a subsonic wall.
    For $\alpha_+ < \alpha_+^{\text{min}}$ or $\alpha_+ > \alpha_+^{\text{max}}$, weak subsonic deflagration solutions are not possible. 
    
    For strongly supercooled transitions, detonation solutions are possible \cite{GYULASSY1984477, PhysRevD.45.3415}. Here, the fluid ahead of the bubble is undisturbed ($\tilde{v}_+=0$), so $v_+ = \xi_w$. Since the pressure in front of the wall is greater than behind it, fluid flows into the bubble as a rarefaction wave until it smoothly comes to a halt at $\xi=c_{s,-}$. Given the convention $v(\xi) \geq 0$ for direct transitions, this implies $\partial_{\xi} v \geq 0$ across the interval $c_{s,-} \leq \xi \leq \xi_w$. From \eqref{eq:eom-2a}, we must therefore have
    \begin{equation}
        \frac{\mu^2(\xi,v)}{c_s^2(T)} - 1 \geq 0.
        \label{eq:detonation-condition}
    \end{equation}
    At the bubble wall, $\mu(\xi=\xi_w, \tilde{v}_-) = v_-$, so for \eqref{eq:detonation-condition} to be satisfied, we require $v_- \geq c_{s,-}$. This is only possible for weak supersonic and Chapman-Jouguet detonations. Furthermore, strong detonations are prohibited as they cannot satisfy the boundary conditions at the wall and the centre of the bubble \cite{steinhardt_relativistic_1982, laine_bubble_1994, barni_hydrodynamics_2024}. 
    
    We note that for inverse transitions, $v(\xi) \leq 0$, so this condition becomes $v_- \leq c_{s,-}$, permitting only weak subsonic and Chapman-Jouguet inverse detonations. The minimum speed of the fluid in front of the wall is determined by the Chapman-Jouguet condition, $v_-=c_{s,-}$ ($v_+=c_{s,+}$ for inverse transitions). In general, this can be calculated by solving the matching conditions at the wall for $v_+$. For the $\mu\nu$ equation of state, we define
    \begin{align}
        v_J(\alpha_+) &= v_+(|v_-|=c_{s,-},\alpha_+) \nonumber \\
        &= \frac{c_{s,-}}{1+3c_{s,-}^2 \alpha_+} \left( 1 \pm \sqrt{3\alpha_+ (1+3c_{s,-}^2\alpha_+ - c_{s,-}^2)} \right).
        \label{eq:vJ}
    \end{align}
    The positive root is called the Jouguet detonation velocity, $v_J^{\text{det}}$, and implies the wall is supersonic ($\xi_w > v_J^{\text{det}} > c_{s,+}$). The thermodynamic properties of the fluid beyond the wall are just those of the symmetric phase, so $T_+ = T_N$ and $w_+=w_N$. From here on, we refer to weak subsonic deflagrations and weak supersonic detonations as just deflagrations and detonations, unless otherwise stated.
    
\begin{figure}[t!]
        \centering
        \includegraphics[width=\textwidth]{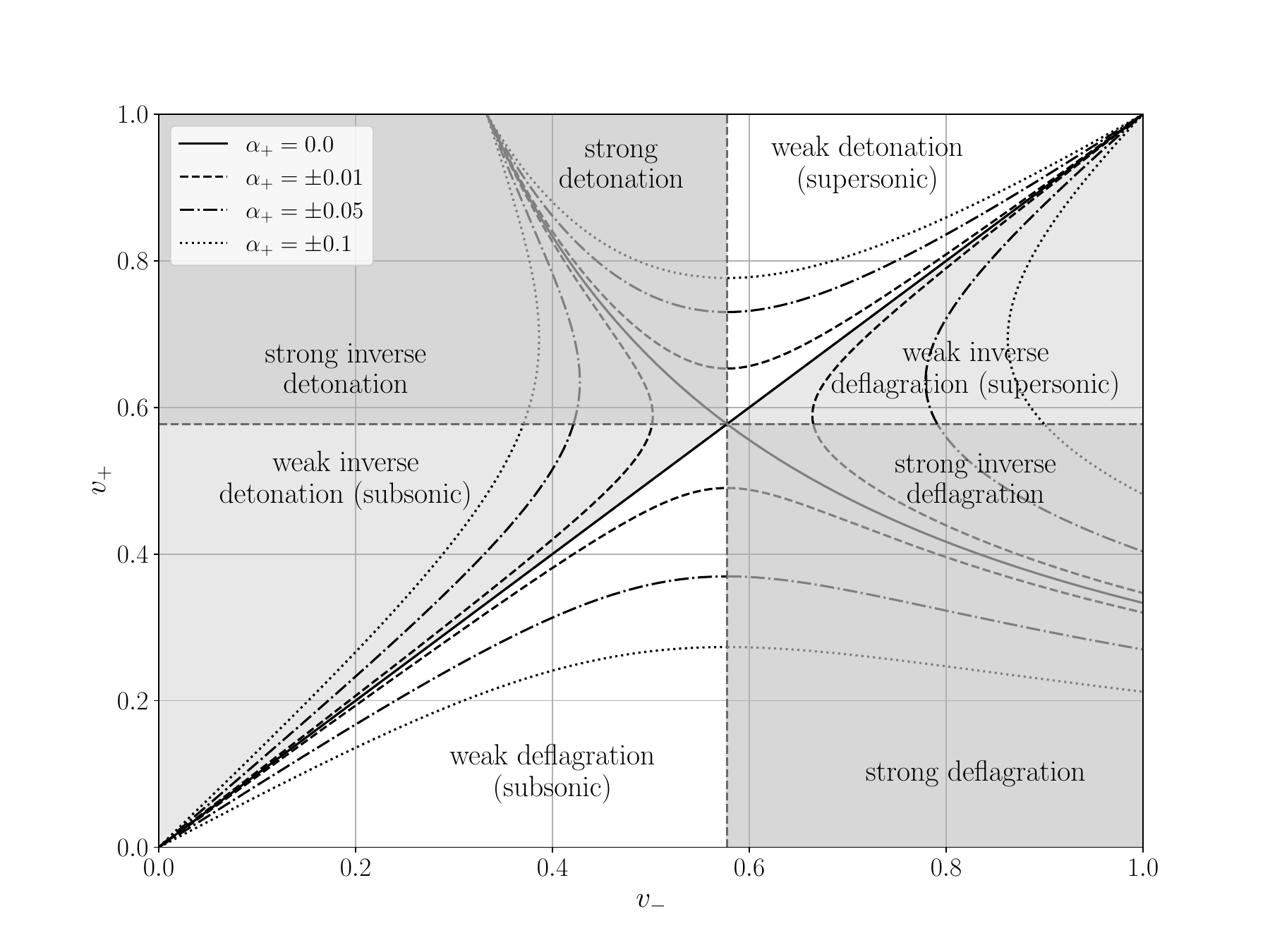}
        \caption{Solution space of hydrodynamic modes during a first-order cosmological phase transition using the bag equation of state (adapted from Figure 9 in \cite{hindmarsh_gravitational_2019}). The curves for different phase transition strength parameters are given by \eqref{eq:vp-from-matching} and constraints for each type of solution are given in \Cref{tbl:hydrodynamic-sols}. Only supersonic weak detonations (detonations), subsonic weak deflagrations (deflagrations), subsonic weak detonations (inverse detonations) and supersonic weak deflagrations (inverse deflagrations) are physically permitted. Inverse transitions correspond to those with $\alpha_+ < 0$.}
        \label{fig:bag-sol-space}
    \end{figure}

    So far, we have constructed consistent solutions for bubble wall velocities $\xi_w < c_{s,+}$ (weak subsonic deflagrations) and $\xi_w > v_J^{\text{det}} > c_{s,+}$ (weak supersonic detonations), which are shown in \Cref{fig:bag-sol-space}. In fact, another stable solution can be constructed for bubble walls that are supersonic and less than the Jouguet detonation velocity. Hydrodynamic simulations \cite{kurki-suonio_supersonic_1995} suggest that strong deflagrations, which we previously mentioned were unstable, tend to decay to form a rarefaction wave similar to a detonation. This gives rise to a type of weak supersonic deflagration called a hybrid. The rarefaction wave causes the fluid behind the bubble wall to be disturbed, so these differ from usual deflagration solutions and are more appropriately described as a superposition of a Chapman-Jouguet deflagration and weak supersonic detonation \cite{laine_bubble_1994} with $v_-=c_{s,-}$ and $v_+ < c_{s,+}$, whose bubble wall velocity satisfies $c_{s,+} \leq \xi_w \leq v_J^{\text{det}}$.
    
\section{Outline of the fluid profile solver}
\label{sec:fp-solver}

\begin{figure}[t!]
    \centering
    \includegraphics[width=\linewidth]{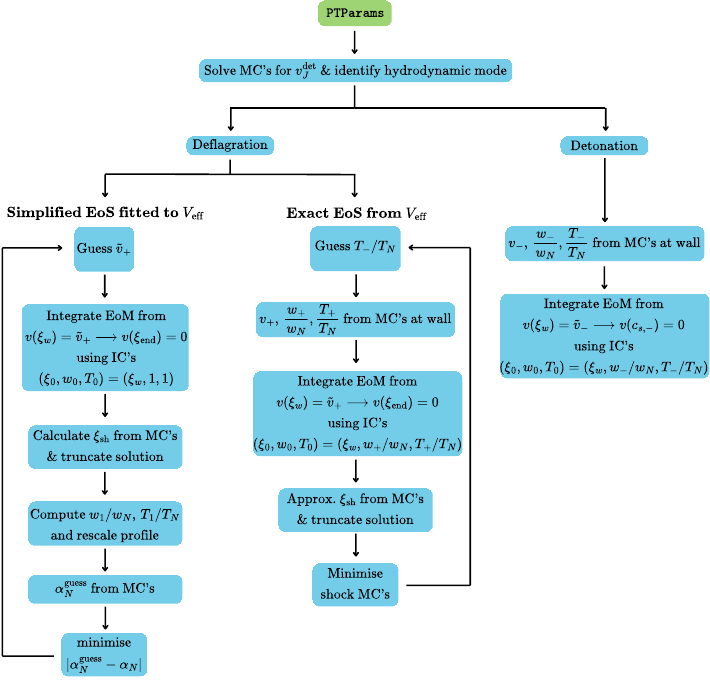}
    \caption{Flow chart describing how fluid profiles are constructed in the \code{FluidProfile} class. Here, `MC' denotes the `matching conditions', `IC' denotes the `initial conditions' and `EoM' is the fluid `equation of motion'.}
    \label{fig:fluid_flow_chart}
\end{figure}

In this section, we provide an overview of how to construct fluid profiles using the \code{HydroGrav} codebase, as well as a detailed description of how the relativistic hydrodynamics part of the calculation works for both simplified and exact equations of state. A graphical description of the construction of fluid profiles is shown in \Cref{fig:fluid_flow_chart}. In principle, any new-physics model that couples to the Standard Model of particle physics can be passed through \code{HydroGrav} to generate fluid profiles and gravitational wave spectra, provided one can accurately determine the effective potential of such a model. 

Fluid profiles are constructed within the \code{FluidProfile} class. First, the hydrodynamic mode is determined using the wall velocity, the speed of sound, and the strength parameter. Deflagration modes have a subsonic wall velocity $\xi_w < c_{s,-}$, hybrid modes have $c_{s,-} \leq \xi_w \leq v_J^{\text{det}}$, and detonation modes have $\xi_w > v_J^{\text{det}}$. Then, the fluid equations of motion \eqref{eq:eom-2} are solved using a fourth-order Runge-Kutta (RK4) method. To avoid numerical instability at $\mu^2(\xi,v)=c_s^2(T)$, the equations of motion we solve are reparametrized as
\begin{subequations}
    \begin{align}
        \frac{\partial \xi}{\partial v} &= \gamma^2 (1-\xi v) \left[ \frac{\mu^2(\xi,v)}{c_s^2(T)} - 1 \right] \frac{\xi}{2v}, \\
        \frac{\partial w}{\partial v} &= \left[ 1 + \frac{1}{c_s^2(T)} \right] w \mu(\xi,v) \gamma^2, \\
        \frac{\partial T}{\partial v} &= T \gamma^2 \mu(\xi,v) ,
    \end{align}
    \label{eq:eom-reparametrised}
\end{subequations}
and integrated across the region where the fluid velocity is non-zero. The methods used to construct the initial conditions, solve the equations of motion, and enforce the matching conditions vary considerably between the simplified (bag and $\mu\nu$) and the exact equation of state from the effective potential. The following two subsections give an overview of the methods used in each case. Unless otherwise stated, we use a subscript `$N$' to denote quantities at the nucleation temperature.
\subsection{Simplified equations of state}
The bag model is just a special case of the $\mu\nu$ model with $c_{s,+}^2=c_{s,-}^2=1/3$, so the methods used to construct the fluid profiles in either case are identical. We first initialise a \newline \code{PTParams\_Bag} object that creates a class of phase transition parameters: the wall velocity $\xi_w$, nucleation temperature $T_N$, strength parameter at nucleation temperature $\alpha_N=\alpha_{\bar{\theta}}(T_N)$, inverse phase transition duration $\beta$, mean bubble separation $R_*$, nucleation type (exponential or simultaneous), the universe constants class, and the speed of sound in the broken and symmetric phases $c_{s,\pm}$ (for the bag model, $c_{s,\pm}^2=1/3$). 

The hydrodynamic mode is simply determined by computing the Jouguet velocity \eqref{eq:vJ}. The solver algorithm for the simplified equations of state is implemented in the \newline \code{solve\_profile} function. For deflagration solutions, the fluid is at rest inside the bubble and beyond the shock, so we construct a solution across $\xi_w < \xi < \xi_{\text{sh}}$. Given some initial condition $v(\xi_w)=\tilde{v}_+$, we solve the equations of motion, initially using $w_+=w_N$ and $T_+=T_N$. We integrate from $v(\xi_w)=\tilde{v}_+$ to $v(\xi_{\text{end}})=0$, where $\xi_{\text{end}}$ is some arbitrary end-point of the integration. The shock front exists at some $\xi_{\text{sh}} \in (c_{s,+}, \xi_{\text{end}})$, which is found by iteratively stepping through the solution until the first shock matching condition in \eqref{eq:shock-matching}, with $v_2=\xi_{\text{sh}}$, is satisfied. The solution is then truncated to the position of the shock to obtain the profiles across the interval $\xi_w < \xi < \xi_{\text{sh}}$. To obtain correctly scaled enthalpy and temperature profiles, we compute $w_1/w_N$ and $T_1/T_N$ from $\xi_\text{sh}$ and normalise the solution such that $w(\xi_{\text{sh}})=w_1/w_N$ and $T(\xi_{\text{sh}})=T_1/T_N$. The enthalpy of the fluid behind the shock is
\begin{equation}
    \frac{w_1}{w_N} = \frac{\xi_{\text{sh}}^2 - c_{s,+}^4}{c_{s,+}^2(1-\xi_{\text{sh}}^2)},
    \label{eq:w1wN-from-shock}
\end{equation}
from the first matching condition in \eqref{eq:matching-conds-sh}. From \eqref{eq:improved-bag-eos-sh}, the temperature behind the shock is simply
\begin{equation}
    \frac{T_1}{T_N} = \left( \frac{w_1}{w_N} \right)^{1/4}.
    \label{eq:T1TN-from-w1wN}
\end{equation}
To determine the initial condition $v(\xi_w)=\tilde{v}_+$ for deflagrations, we construct a residual function \code{alN\_residual} that computes the strength parameter $\alpha_N$ from the matching conditions at the wall and compares it to the input value of $\alpha_N$. To compute $\alpha_N$ using our `guess' $\tilde{v}_+$, we first solve the fluid equations of motion with the guess value to obtain $w_+/w_N$, the enthalpy in front of the wall. We then compute $\alpha_+$ from the matching condition at the wall by inverting \eqref{eq:vp-from-matching},
    \begin{equation}
        \alpha_+ = \frac{\gamma_+^2}{3} \left( \frac{v_+^2}{c_{s,-}^2} - \frac{v_+ v_-}{c_{s,-}^2} - \frac{v_+}{v_-} + 1 \right),
    \end{equation}
where $v_-=\xi_w$ for deflagrations. The definition of the strength parameter \eqref{eq:alpha-munu} gives us the relation
\begin{equation}
    \alpha_N = \frac{w_+}{w_N} \alpha_+ + \frac{1}{3} \frac{c_{s,+}^2/c_{s,-}^2 - 1}{1 + c_{s,+}^2} \left( \frac{w_+}{w_N} + 1 \right),
\end{equation}
which we use to determine $\alpha_N^{\text{guess}}$ and then calculate the corresponding residual $|\alpha_N^{\text{guess}} - \alpha_N|$. The residual is computed in a bracket $(\tilde{v}_+^{\text{min}},\tilde{v}_+^{\text{max}})$ (most straightforwardly taken to be $(0,1)$) and minimised using a golden section minimisation method to find the correct $\tilde{v}_+$. In practice, the residual only exists in a neighbourhood of the true value for $\tilde{v}_+$. Outside of this neighbourhood, there are no solutions that satisfy both of the matching conditions at the bubble wall and shock front. 

The energy density fluctuation profile $\lambda(\xi)$, defined as
\begin{equation}
    \lambda(\xi) = \frac{e(\xi) - e_N}{w_N}, \label{eq:la-prof}
\end{equation}
is also computed.  This quantity will be used in \eqref{eq:l-integral} in the sound shell model.  Here, $e(\xi) = e_+(T(\xi))$ for deflagrations. Using the $\mu\nu$ equation of state, we note that the relation \eqref{eq:T1TN-from-w1wN} holds across the entire deflagration profile. That is, $T(\xi)/T_N = (w(\xi)/w_N)^{1/4}$. The energy density fluctuation profile is therefore
\begin{equation}
    \lambda^{\text{def}}(\xi) = \frac{w(\xi)/w_N - 1}{1+c_{s,+}^2}.
\end{equation}
Detonation solutions are considerably simpler. The fluid is at rest beyond the wall and comes to a halt inside the bubble at $\xi=c_{s,-}$, so we construct a solution across $c_{s,-} < \xi < \xi_w$. We use the initial conditions $(\xi_0,w_0,T_0)=(\xi_w, w_-/w_N, T_-/T_N)$ and integrate from $v(\xi_w)=\tilde{v}_-$ to $v(c_{s,-})=0$, where $\tilde{v}_-=\mu(\xi_w, |v_-|)$ and $v_-$ is determined from inverting the matching condition \eqref{eq:vp-from-matching},
\begin{align}
    v_- = \frac{1}{2} \left[\vphantom{\rule{0pt}{5.4ex}}\right.& \left( v_+ + \frac{c_{s,-}^2}{v_+} \left(1 - 3\alpha_+ (1-v_+^2)\right) \right) \nonumber \\ &\pm \sqrt{\left( v_+ + \frac{c_{s,-}^2}{v_+} \left(1 - 3\alpha_+ (1-v_+^2)\right) \right)^2 - 4c_{s,-}^2} \left.\vphantom{\rule{0pt}{5.4ex}}\right],
\end{align}
where $v_+=\xi_w$ and $\alpha_+=\alpha_N$ for detonations. The enthalpy just behind the wall is determined from the matching condition \eqref{eq:matching-conds}
\begin{equation}
    w_-=\frac{w_+ v_+ \gamma_+^2}{v_- \gamma_-^2}, \label{eq:wm-from-matching}
\end{equation}
with $w_+ = w_N$ for detonations. From the bag equation of state \eqref{eq:bag-eos}, $w_{\pm}=(1+c_{s,\pm}^2) a_{\pm} T_{\pm}^4$. The temperature behind the wall is then
\begin{equation}
    \frac{T_-}{T_N} = \left[ \frac{a_+}{a_-} \left(\frac{1+c_{s,+}^2}{1+c_{s,-}^2} \right) \frac{w_-}{w_N} \right]^{1/4},
\end{equation}
and the ratio $a_+/a_-$ is approximated to $1$ in this work. These initial conditions are then used to integrate the equations of motion \eqref{eq:eom-reparametrised} from $v(\xi_w)=\tilde{v}_-$ to $v(c_{s,_-})=0$. The energy density fluctuation profile for detonations is slightly more complex with $e(\xi) = e_-(T(\xi))$, since the bag constant $\epsilon$ remains in the numerator of \eqref{eq:la-prof}. To express $\lambda(\xi)$ in terms of the enthalpy, sound speed, and strength parameter, we solve \eqref{eq:alpha-munu} for $\epsilon$, and we find
\begin{equation}
    \lambda^{\text{det}}(\xi) = \frac{w(\xi)/w_N - 1 - 3c_{s,-}^2 \alpha_N}{1+c_{s,-}^2}.
\end{equation}
Hybrid solutions require us to stitch together both a deflagration solution ($\xi_w < \xi < \xi_{\text{sh}}$) and a detonation solution ($c_{s,-} < \xi < \xi_w$). We do this using the usual deflagration/detonation solvers described above, but with the boundary condition $v_-=c_{s,-}$ in the deflagration part of the profiles and using the matching conditions at the wall \eqref{eq:matching-conds} to obtain the initial condition for the detonation part of the profiles.

\subsection{Exact equation of state}
To construct the fluid profiles directly from the effective potential, we first initialise a \code{PTParams\_Veff} object which takes in the same phase transition parameters as the simplified equations of state, except for the speed of sound in each phase. Instead, an \code{EquationOfState} object is passed in, which contains the pressure $p$ and energy density $e$ in each phase as a function of temperature from some effective potential. The equation of state data is stored in the class, and the speed of sound $c_s^2(T)=\mathrm{d}p/\mathrm{d}e$ in each phase is evaluated numerically. 

The \code{PTParams\_Veff} object is then used to initialise the \code{FluidProfile} class. Constructing fluid profiles for the exact equation of state is more involved than for simplified equations of state, since the algebraic forms of the thermodynamic quantities $p(T)$, $e(T)$, $w(T)$, and $s(T)$ are unknown. Consequently, the matching conditions in equation~\eqref{eq:matching-conds-2} must be imposed numerically using a root-finding algorithm. Throughout most of the code, we use a bounded 2D Newton's method solver to find the root of
\begin{equation}
    f_1^{\text{w}} = v_- v_+ - \frac{p_- - p_+}{e_- - e_+}, \quad f_2^{\text{w}} = \frac{v_-}{v_+} - \frac{e_+ + p_-}{e_- + p_+},
\end{equation}
for matching at the wall and
\begin{equation}
    f_1^{\text{sh}} = v_1 v_2 - \frac{p_1 - p_2}{e_1 - e_2}, \quad f_2^{\text{sh}}  = \frac{v_1}{v_2} - \frac{e_2 + p_1}{e_1 + p_2},
    \label{eq:matching-shock-veff}
\end{equation}
for matching at the shock. Here, $p_\pm$ and $e_\pm$ are defined in \eqref{eq:EoS-exact}, and $p_{1,2}$ and $e_{1,2}$ are defined similarly in \eqref{eq:EoS-exact-shock}. 

Determining the hydrodynamic mode requires us to compute the Jouguet detonation velocity, which is given by the fluid velocity in front of the bubble wall when $|v_-|=c_{s,-}$. We employ the root-finding algorithm to calculate $v_+=v_J^{\text{det}}$ and $T_-$ from $v_-$ and $T_+=T_N$ (although only the former is needed). The profiles are then constructed using the \code{FluidProfile::solve\_profile\_veff} function.

For deflagration solutions, we utilise the same method to find the shock front $\xi_{\text{sh}}$ as in the simplified equation of state case, with the exception of minimising \eqref{eq:matching-shock-veff} to enforce the matching condition at the shock. The slight caveat in determining the initial conditions is that the phase transition strength parameter $\alpha_{\theta}$ is not defined for a generic equation of state. Instead, we construct a different residual. Namely, we start with a `guess' value for the temperature behind the wall $T_-$ and use the matching conditions at the wall to obtain $v_+$, $w_+$, and $T_+$. We use this as the initial condition to evolve the equations of motion and then truncate the solution at the shock front, where the first shock matching condition $f^{\text{sh}}_1$ is minimised. This routine gives us $v_1$, the velocity of the fluid just behind the wall. Once again, we use a golden section minimisation method across the interval $(T_-^{\text{min}}, T_-^{\text{max}})$, defined by the minimum and maximum temperature values passed in from the equation of state, to find $v_1$ that minimises the second shock matching equation $f^{\text{sh}}_2$. We note that a fallback method is used in case the shock matching conditions do not converge to a sufficient level of precision, which can occur for extremely narrow fluid profiles that form when $|\xi_{\text{sh}} - \xi_w| \ll 1$. The fallback method instead enforces the shock matching conditions of the $\mu\nu$ model \eqref{eq:shock-matching}, using $c_{s,+/-}=c_{s,s/b}(T_{+/-})$. 

For detonation solutions, the solver works similarly to the simplified equation of state case, except we must modify the method used to construct the initial conditions $(\xi_0, w_0, T_0)=(\xi_w,w_-/w_N,T_-/T_N)$. We use the root-finding algorithm to determine $v_-$ and $T_-/T_N$ from $v_+=\xi_w$ and $T_+/T_N=1$. The enthalpy behind the wall is found using \eqref{eq:wm-from-matching}, and then the equations of motion are integrated from $v(\xi_w)=\tilde{v}_-$ to $v(c_{s,_-})=0$. 

The energy density fluctuation profile for both types of solutions is simply determined from the energy density, $e_{s,b}(T)$, and the temperature profile, $T(\xi)$, where we use the energy density in the symmetric ($s$) phase for deflagrations and in the broken ($b$) phase for detonations. Hybrid solutions are constructed in the same manner as before, using the condition $v_-=c_{s,b}(T=T_-)$ as the endpoint of integration, where $c_{s,b}(T)$ is the speed of sound in the broken phase.

To construct a gravitational wave spectrum for any equation of state, the phase transition parameters (\code{PTParams\_Bag} or \code{PTParams\_Veff}) are passed directly into the \code{GWSpec} function.

% Solver validation
\subsection{Validation of numerical methods}
Unit tests are provided to test the correct functionality of the construction of fluid profiles (\code{test\_fluidprofile.cpp}), the construction of sound shell model spectra (\code{test\_powerspec.cpp}), and the detector noise spectrum calculations (\code{test\_detector.cpp}). The unit test \code{test\_maths.cpp} is also provided for testing the convergence of all the numerical routines used in \code{HydroGrav}:
\begin{itemize}
    \item \textit{Golden section minimisation:} Minimising residuals to obtain $v_+$ (simplified equations of state) and $T_-$ (exact equation of state).
    \item \textit{Bounded 2D Newton's method solver:} Determining the matching conditions at the wall and the shock.
    \item \textit{RK4 solver:} Integrating the fluid equation of motion.
\end{itemize}
The golden section minimisation routine is tested against several functions whose minimum is known analytically, and converges to within $10^{-8}$ of their true value. The tolerance on the minimiser can be manually adjusted from its default value of $10^{-8}$, which is used in the construction of fluid profiles, when calling it.

The 2D Newton solver is also tested against several systems of equations whose solutions are known analytically, and uses a default tolerance of $10^{-8}$ on the norm of the two residual functions. However, this does not guarantee the convergence of each residual function individually, so we further check that each residual is smaller than $10^{-4}$ when solving the matching conditions. We note that for the exact equation of state, the residual functions minimised at the wall and shock to enforce the matching conditions are scaled by $s=\mathrm{max}(p_\pm, e_\pm)$ to obtain $\mathcal{O}(1)$ residuals, which considerably improves the success rate of the solver. In this case, the tolerance is set to $10^{-4} \times s$ to ensure the unscaled residual functions converge.

The RK4 solver is tested for two ordinary differential equations whose solutions are a simple harmonic oscillator and an exponential function. Both numerical solutions converge to within $10^{-10}$ of their analytic values. By default, $N=5000$ steps are used to integrate the fluid profiles using the RK4 solver, which can be adjusted as an input to the \code{FluidProfile} constructor. We find that the numerical values of the fluid velocity, enthalpy, and temperature on either side of the wall/shock, as well as the position of the shock, converge to at least $0.01\%$ of the values obtained when using $N=50000$ steps.

% GWs from the SSM
\section{Gravitational waves from a first-order phase transition}
\label{sec:GWs}
Gravitational waves are described by perturbations of the metric tensor in linearised gravity. We consider an expanding Universe, described by the Friedmann–Lema\^{i}tre–Robertson–Walker (FLRW) metric,
\begin{equation}
    \mathrm{d}s^2 = a^2(\tau) \left[ -\mathrm{d}\tau^2 + (\delta_{ij} + h_{ij}) \mathrm{d}x^i \mathrm{d}x^j \right],
\end{equation}
where $a$ is the scale factor and $\eta^{ij} h_{ij} = \partial^i h_{ij} = 0$ in the transverse-traceless (TT) gauge, with $\eta^{ij}$ being the Minkowski metric. Defining $l_{ij} = a h_{ij}$, Einstein's equations in momentum space read
\begin{equation}
    (\partial_{\tau}^2 + k^2) \tilde{l}_{ij}(\tau,\bm{k}) = \frac{6 \mathcal{H}_*}{\tau} \Pi_{ij}(\tau, \bm{k}). \label{eq:gw-EEs}
\end{equation}
Here, $\Pi_{ij}$ is the transverse-traceless projection of the stress-energy, $T_{ij}$, that sources the gravitational waves. For a first-order phase transition, this is given by \eqref{eq:FOPT-EM-total}. We define the conformal Hubble parameter, $\mathcal{H}=a'/a$, where a prime denotes a derivative with respect to the conformal time, $\tau$. We let $\tau_*$ denote the time when gravitational waves start being produced, with $a(\tau_*)=1$ such that $a(\tau)=\mathcal{H}_* \tau$, where $\mathcal{H}_*=\mathcal{H}(\tau_*)$. We further define the projection operator
\begin{equation}
    \Lambda_{ij,lm} = P_{il} P_{jm} - \frac{1}{2} P_{ij} P_{lm},
\end{equation}
where $P_{ij} = \delta_{ij} - \hat{k}_i \hat{k}_j$, such that
\begin{equation}
    \bar{e} \Pi_{ij}(\tau, \vec{k}) = \Lambda_{ij,lm} T^{lm}(\tau, \bm{k}).
\end{equation}
Here, $\bar{e} = 3\mathcal{H}^2/(8\pi G a^2)$ is the average energy density in the Universe, where $G$ is Newton's gravitational constant. For gravitational waves sourced over the duration $\delta \tau = \tau_{\text{fin}} - \tau_*$, the solutions to \eqref{eq:gw-EEs} are
\begin{equation}
    \tilde{l}_{ij} (\tau, \bm{k}) = 6\mathcal{H}_*
    \begin{dcases}
        \int_{\tau_*}^{\tau} \frac{\mathrm{d}\tau_1}{\tau_1} \Pi_{ij}(\tau_1,\bm{k}) \frac{\sin{[k(\tau-\tau_1)]}}{k}, & \tau \in [\tau_*, \tau_{\text{fin}}] \\
        \int_{\tau_*}^{\tau_{\text{fin}}} \frac{\mathrm{d}\tau_1}{\tau_1} \Pi_{ij}(\tau_1,\bm{k}) \frac{\sin{[k(\tau-\tau_1)]}}{k}, & \tau > \tau_{\text{fin}} \\
    \end{dcases}.
\end{equation}
The gravitational wave power spectrum is defined as
\begin{equation}
    \Omega_{\text{GW}}(k) = \frac{1}{\bar{e}} \frac{\mathrm{d} e_{\text{GW}}}{\mathrm{d} \ln{k}}, \label{eq:gw-power-spec}
\end{equation}
where the energy density of the gravitational waves is given by
\begin{equation}
    e_{\text{GW}} = \frac{1}{32 \pi G} \langle h_{ij}^{'}(\tau,\bm{x}) h^{'ij}(\tau,\bm{x}) \rangle \approx \frac{1}{12 \mathcal{H}^2 a^2} \langle l_{ij}^{'}(\tau,\bm{x}) l^{'ij}(\tau,\bm{x}) \rangle.
\end{equation}
To determine \eqref{eq:gw-power-spec}, we must evaluate the two-point function of the gravitational field:
\begin{align}
    \langle \tilde{l}_{ij}^{*'}(\tau_1, \bm{k}) \tilde{l}^{'ij}(\tau_2, \bm{k}_2 \rangle = (6\mathcal{H}_*)^2 \int_{\tau_*}^{\tau_{\text{fin}}} \frac{\mathrm{d}\tau_1}{\tau_1} \int_{\tau_*}^{\tau_{\text{fin}}} \frac{\mathrm{d}\tau_2}{\tau_2} &\cos{[k(\tau-\tau_1)]} \cos{[k_2(\tau-\tau_2)]} \nonumber \\ &\times\langle \Pi_{ij}(\tau_1,\bm{k}) \Pi^{ij}(\tau_2,\bm{k}_2) \rangle. \label{eq:lij-2pt-func}
\end{align}
The two-point function of the stress-energy tensor that appears in \eqref{eq:lij-2pt-func} defines the unequal-time correlator (UETC) of anisotropic stress, $E_{\Pi}$, which is given by
\begin{equation}
    \langle \Pi_{ij}(\tau_1,\bm{k}) \Pi^{ij}(\tau_2,\bm{k}_2) \rangle = (2\pi)^6 \delta^3(\bm{k} - \bm{k}_2) \frac{E_{\Pi}(\tau_1, \tau_2, k)}{4\pi k^2},
    \label{eq:gw-UETC}
\end{equation}
where $k=|\bm{k}|$. Averaging over highly oscillatory modes, the gravitational wave spectrum today is \cite{pol_characterization_2024}
\begin{equation}
    \Omega_{\text{GW}}(k) \approx \frac{3k}{2} \mathcal{T}_{\text{GW}} \int_{\tau_*}^{\tau_{\text{fin}}} \frac{\mathrm{d}\tau_1}{\tau_1} \int_{\tau_*}^{\tau_{\text{fin}}} \frac{\mathrm{d}\tau_2}{\tau_2} E_{\Pi}(\tau_1, \tau_2, k) \cos{k(\tau_1-\tau_2)}. \label{eq:gw-spec}
\end{equation}
Here, $\mathcal{T}_{\text{GW}}$ is the transfer function, which describes the gravitational redshift of the spectrum from the time gravitational waves were sourced to today, and it is defined by
\begin{equation}
    h^2\mathcal{T}_{\text{GW}} = \left( \frac{a_*}{a_0} \right)^4 \left( \frac{H_*}{H_0/h} \right)^2 \approx 1.6 \times 10^{-5} \left( \frac{100}{g_*} \right)^{1/3},
\end{equation}
where $H_*$ is the Hubble rate and $g_*$ is the number of degrees of freedom at $\tau_*$, and the Hubble rate today is $H_0$, with $h= H_0 / 100 \text{ km/s/Mpc}$. Unless otherwise stated, we use `$*$' and `$0$' subscripts to denote quantities at the start of gravitational wave production and today, respectively. 

Determining the gravitational wave spectrum amounts to constructing an appropriate stress-energy tensor and evaluating the UETC for any given gravitational wave source. For a first-order phase transition, there are three primary mechanisms for producing gravitational waves \cite{athron_cosmological_2024}. 1) Bubble collisions: originally assumed to be spherically symmetric, they collide with each other as the transition progresses, breaking their spherical symmetry and producing quadrupole and higher-order moments in the gravitational field that source gravitational waves. 2) Sound waves: the energy carried by the bubble wall produces sound waves in the surrounding fluid, which subsequently interact and source gravitational waves. 3) Turbulence: non-linear fluid dynamics, including turbulence in the fluid, further contribute to multipole moments that produce gravitational waves.

Numerical simulations \cite{hindmarsh_gravitational_2014, hindmarsh_numerical_2015} indicate gravitational waves sourced by sound waves provide the dominant contribution, while those generated by bubble collisions and turbulence are sub-dominant for thermal phase transitions without significant supercooling. Fitting formulas, such as those described in \cite{hindmarsh_numerical_2015, Hindmarsh:2017gnf, Caprini:2019egz}, capture the contribution to the gravitational wave spectrum from each of these sources by numerically estimating the UETC. The sound wave contribution is often described by a single broken power law; for instance \cite{Hindmarsh:2017gnf}
\begin{equation}
    h^2 \Omega_{\text{GW}}^{\text{fit}}(f) = 2.061 h^2 \left(\frac{\bar{w}}{\bar{e}}\right)^2 F_{\text{gw},0} \bar{U}^4_f S_{\text{sw}}(f) \tilde{\Omega}_{\text{gw}} \min(H_* R_* / \bar{U}_f, 1) (H_* R_*),
    \label{eq:sw-fit}
\end{equation}
where
\begin{align}
    F_{\mathrm{gw},0} &= 3.57 \times 10^{-5} \left(\frac{100}{g_*}\right)^{\frac{1}{3}}, \\
    S_{\mathrm{sw}}(f) &= \left(\frac{f}{f_{\mathrm{sw}}}\right)^3 \left(\frac{7}{4 + 3\left(f/f_{\mathrm{sw}}\right)^2}\right)^{\frac{7}{2}}, \\
    \frac{f_{\mathrm{sw}}}{1\,\mu\mathrm{Hz}} &= 2.6 \left(\frac{z_p}{10}\right) \left(\frac{T_*}{100\,\mathrm{GeV}}\right) \left(\frac{g_*}{100}\right)^{\frac{1}{6}} \left(\frac{1}{H_* R_*}\right),
\end{align}
and $z_p \sim 10$ and $\tilde{\Omega}_{\text{gw}} \sim 0.012$ from numerical simulations. 
The root-mean-square fluid velocity~\cite{Hindmarsh:2017gnf}, $\bar{U}_f$, defined as
\begin{equation}
    \bar{U}_f^2 = \frac{1}{\bar{w}\mathcal{V}}\int_{\mathcal{V}}d^3x T_{ii}^f,
\end{equation}
where $\mathcal{V}$ is the averaging volume.
In practice, it can be approximated in terms of the efficiency factor, $\kappa$, which describes how much of the vacuum energy is converted into the bulk kinetic energy of the fluid.
Assuming the bag equation of state~\cite{Hindmarsh:2017gnf, Caprini:2019egz}, one has
\begin{equation}
     \frac{\bar{w}}{\bar{e}} \bar{U}_f^2 = \frac{\kappa \alpha_N}{1+\alpha_N},
\end{equation}
where \cite{espinosa_energy_2010}
\begin{equation}
    \kappa = \frac{3}{\epsilon \xi_w^3} \int \mathrm{d}\xi \, \xi^2 w(\xi) v(\xi)^2 \gamma^2.
\end{equation}
In contrast, the sound shell model \cite{hindmarsh_sound_2018, hindmarsh_gravitational_2019, Guo:2020grp, Wang:2021dwl} is a semi-analytic approach for estimating the gravitational wave spectrum from sound waves, offering a more detailed description of the spectral shape than traditional fitting formulas. In the sound shell model, the UETC is proportional to the four-point function of the fluid velocity field and incorporates two characteristic scales: the mean bubble separation and thickness of the sound shells. In the next section, we provide a brief overview of the sound shell model, following the notation and conventions used in \cite{pol_characterization_2024, Tian:2025zlo}, and then compare the gravitational wave spectra obtained from the simplified and exact equations of state in section \ref{sec:gw-comparison}.

\section{Outline of the sound shell model}
\label{sec:ssm}
The total stress-energy \eqref{eq:FOPT-EM-total} includes a full non-linear description of the fluid and external field(s) that drive the phase transition. In the sound shell model, we assume the bubble separation is much less than the Hubble distance, $1/H_*$, at the transition and consider only the contributions from linearised fluid motion to the gravitational wave spectrum. Moreover, the velocity of the fluid is assumed to be non-relativistic, such that the Lorentz factor is $\gamma \sim 1$ and the relevant part of the stress-energy is
\begin{equation}
    T_{ij}(\tau, \bm{k}) = \bar{w} \int \frac{\mathrm{d}^3\bm{p}}{(2\pi)^3} u_i(\tau, \bm{p}) u_j(\tau, \tilde{\bm{p}}),
\end{equation}
where $u_i = \gamma v_i$, $v_i$ is the fluid velocity, $\bar{w}$ is the average enthalpy of the universe, and $\tilde{\bm{p}} = \bm{k} - \bm{p}$. The velocity field of the fluid is further assumed to be Gaussian, so the four-point correlation functions can be expressed as a sum of two-point functions,
\begin{align}
    \langle T_{ij}(\tau_1,\bm{k}) T_{lm}^*(\tau_2,\bm{k}) \rangle = &\bar{w}^2 \int \frac{\mathrm{d}^3\bm{p}_1}{(2\pi)^3} \int \frac{\mathrm{d}^3\bm{p}_2}{(2\pi)^3} \nonumber \\
    & \times \left[ \langle u_i(\tau_1, \bm{p}_1) u_l^*(\tau_2, \bm{p}_2 \rangle \langle u_j(\tau_1, \tilde{\bm{p}}_1) u_m^*(\tau_2, \tilde{\bm{p}}_2 \rangle \right. \nonumber \\
    &\left. \quad + \langle u_i(\tau_1, \bm{p}_1) u_m^*(\tau_2, \bm{p}_2 \rangle \langle u_j(\tau_1, \tilde{\bm{p}}_1) u_l^*(\tau_2, \tilde{\bm{p}}_2 \rangle \right].
\end{align}
For a homogeneous, isotropic and irrotational fluid, the two-point function of the velocity field is proportional to a spectral function, $E_{\text{kin}}(\tau_1, \tau_2, k)$, defined by
\begin{equation}
    \langle u_i(\tau_1, \bm{k}) u_j^*(\tau_2, \bm{k}_2) \rangle = (2\pi)^6 \hat{k}_i \hat{k}_j \delta^3(\bm{k}-\bm{k}_2) \frac{2 E_{\text{kin}}(\tau_1, \tau_2, k)}{4\pi k^2}, \label{eq:v-2pt-func}
\end{equation}
where $\hat{k}_i = k_i/k$ and $k=|\bm{k}|$. Combining \eqref{eq:v-2pt-func} with the definition \eqref{eq:gw-UETC}, the UETC is given by
\begin{equation}
    E_{\Pi}(\tau_1, \tau_2, k) = 2k^2 \bar{w}^2 \int_{-1}^1 \mathrm{d}z \int_0^{\infty} \mathrm{d}p \, \frac{p^2}{\tilde{p}^4} (1-z^2)^2 E_{\text{kin}}(\tau_1, \tau_2, p) E_{\text{kin}}(\tau_1, \tau_2, \tilde{p}),
    \label{eq:UETC-Ekin}
\end{equation}
where $z =\hat{\bm{k}} \cdot \hat{\bm{p}}$. To evaluate the spectral function, we define the energy density fluctuation, $\lambda = (e - \bar{e})/\bar{w}$, and construct solutions to the linearised fluid equations
\begin{subequations}
\begin{align}
    &u_i'(\tau,\bm{k}) - i k_i c_s^2 \lambda(\tau, \bm{k}) = 0, \\
    &\lambda'(\tau,\bm{k}) - i k_i u^i(\tau, \bm{k}) = 0,
\end{align}
\end{subequations}
where $c_s^2 = \mathrm{d}\bar{p}/\mathrm{d}\bar{e}$ is the average speed of sound in the symmetric phase. Solutions can be constructed by defining the longitudinal velocity field
\begin{equation}
    u(\tau, \bm{k}) = \sum_{s=\pm} A_s(\bm{k}) e^{i s k c_s (\tau-\tau_*)}, \label{eq:longitudinal-velocity-field}
\end{equation}
    where $u_i = \hat{k}_i u(\tau, \bm{k})$ is the fluid velocity projected along the direction of its three-momentum. In the sound shell model, the velocity and energy density fields are taken to be a linear superposition of self-similar fluid profiles, each of which describes the dynamics of a single expanding bubble. For $N$ bubbles, the coefficients $A_{\pm}$ are \cite{hindmarsh_gravitational_2014, hindmarsh_gravitational_2019}
\begin{equation}
    A_{\pm} (\bm{k}) = \sum_{n=1}^N \mathcal{A}_{\pm} (\chi) T_n^3 e^{i \bm{k} \cdot \bm{x}_0^{(n)}}, \quad \mathcal{A}_{\pm} (\chi) = -\frac{i}{2} \left[ f'(\chi) \pm i c_s l(\chi) \right],
\end{equation}
where the $n$-th bubble has a lifetime of $T_n = \tau_* - \tau_0^{(n)}$, with $\tau_0^{(n)}$ and $x_0^{(n)}$ being the time and position of nucleation, respectively, and $\chi = k T_n$. The functions $f$ and $l$ are integrals over the single bubble fluid profiles,
\begin{subequations}
    \begin{align}
        f(\chi) &= \frac{4\pi}{\chi} \int_0^{\infty} \mathrm{d}\xi \, v(\xi) \sin(\chi \xi), \label{eq:f-integral} \\
        l(\chi) &= \frac{4\pi}{\chi} \int_0^{\infty} \mathrm{d}\xi \, \xi \lambda(\xi) \sin(\chi \xi), \label{eq:l-integral}
    \end{align}
    \label{eq:profile-integrals}
\end{subequations}
which are found by solving the hydrodynamic equations of motion \eqref{eq:eom-2} and are shown in \Cref{fig:fluid_profiles}\footnote{The energy density fluctuation profile is constructed entirely from the enthalpy profile (or equivalently, the temperature profile), as we show in section \ref{sec:fp-solver}.}. Combining \eqref{eq:longitudinal-velocity-field} with \eqref{eq:v-2pt-func} and taking the limit as $k \gg 1/(2 c_s \delta \tau)$, where $\delta \tau$ is the sound wave duration, the spectral function is \cite{hindmarsh_gravitational_2019}
\begin{equation}
    E_{\text{kin}}(\tau_1, \tau_2, k) \approx E_{\text{kin}}(k) \cos(k c_s \tau_-) , \label{eq:kinetic-spectral-func}
\end{equation}
where $\tau_- = \tau_2 - \tau_1$.  The function $E_{\text{kin}}(k)$ is often referred to as the kinetic (or velocity) spectrum and is given by \cite{pol_characterization_2024}
\begin{equation}
    E_{\text{kin}}(k) = \frac{k^2}{2\pi^2 \beta^6 R_*^3} \int_0^{\infty} \mathrm{d}\tilde{T} \, \nu(\tilde{T}) \tilde{T}^6 |\mathcal{A}_+(\tilde{T}k/\beta)|^2, \label{eq:Ekin}
\end{equation}
where $\beta$ is the inverse duration of the phase transition, $\tilde{T} = T\beta$ is the normalised bubble lifetime, and $R_*=n(t)^{-1/3}$ is the mean bubble separation, with $n(t)$ being the number density of bubbles during the phase transition, which can be approximated as $R_*=(8 \pi)^{1/3} \xi_w/\beta$ \cite{Megevand:2016lpr, PhysRevD.45.3415}. The bubble lifetime distribution, $\nu(\tilde{T})$, depends on the nucleation process and is given by \cite{hindmarsh_gravitational_2019}
\begin{equation}
    \nu_{\text{exp}}(\tilde{T}) = e^{-\tilde{T}}, \quad \nu_{\text{sim}}(\tilde{T}) = \frac{1}{2} \tilde{T}^2 e^{-\frac{1}{6}\tilde{T}^3},
\end{equation}
for exponential and simultaneous nucleation. Combining \eqref{eq:UETC-Ekin} and \eqref{eq:Ekin} with \eqref{eq:gw-spec}, the gravitational wave spectrum due to sound waves, as measured today, is \cite{pol_characterization_2024}
\begin{align}
    \Omega_{\text{GW}}(K) &= 3 K^3 \left( \frac{\bar{w}}{\bar{e}} \right)^2 \left( \frac{\Omega_{\mathcal{K}}}{\mathcal{K}} \right)^2 \mathcal{T}_{\text{GW}} \int_0^{\infty} \mathrm{d}P \int_{-1}^1 \mathrm{d}z \, (1-z^2)^2 \frac{P^2}{\tilde{P}^4} \nonumber \\ &\, \quad \times \zeta_{\text{kin}}(P) \zeta_{\text{kin}}(\tilde{P}) \Delta(\delta \tau, k, p, \tilde{p}),
    \label{eq:gw-ssm}
\end{align}
where we have defined the dimensionless momentum parameter, $K=k R_*$ (and $P=pR_*$), and the normalised kinetic spectrum, $\zeta_{\text{kin}}(K) = E_{\text{kin}}(K) / E_{\text{kin}}^{\text{max}}$. The total kinetic energy density, $\Omega_{\mathcal{K}}$, is defined as\footnote{Although $\zeta_{\text{kin}}$ fails to accurately describe the UETC at small $K$ due to the assumption $k \gg 1/(2c_s\delta \tau)$ made in \eqref{eq:kinetic-spectral-func}, this contributes negligibly to $\mathcal{K}$ \cite{pol_characterization_2024}.}
\begin{align}
    \Omega_{\mathcal{K}} = \int_0^{\infty} \mathrm{d}k \, E_{\text{kin}}(\tau, \tau, k)
    = \frac{E^{\text{max}}_{\text{kin}}}{R_*} \mathcal{K}; \quad \mathcal{K} \approx \int_0^{\infty} \mathrm{d}K \, \zeta_{\text{kin}}(K).
\end{align}
The function $\Delta$ is given by
\begin{equation}
    \Delta(\delta \tau, k, p, \tilde{p}) = \sum_{m,n=\pm1} \Delta_{mn}(\delta \tau, \hat{p}_{mn}),
\end{equation}
for the sound shell model, where $\hat{p}_{mn}=(p+m\tilde{p})c_s+nk$,
\begin{equation}
    \Delta_{mn} (\delta \tau, \hat{p}_{mn}) = \frac{1}{4} \left[ \Delta \mathrm{Ci}^2(\tau_\text{fin}, \hat{p}_{mn}) + \Delta \mathrm{Si}^2(\tau_\text{fin}, \hat{p}_{mn})\right],
\end{equation}
and
\begin{subequations}
    \begin{align}
        \Delta \mathrm{Ci}(\tau, p) &= \mathrm{Ci}(p\tau) - \mathrm{Ci}(p\tau_*), \\
        \Delta \mathrm{Si}(\tau, p) &= \mathrm{Si}(p\tau) - \mathrm{Si}(p\tau_*),
    \end{align}
\end{subequations}
where, $\mathrm{Ci}$ and $\mathrm{Si}$ are the usual cosine and sine integrals, respectively. 

The integrals \eqref{eq:profile-integrals} are highly oscillatory for small $\chi$, so we use a Filon quadrature integrator for $\chi < 10$. For $\chi$ above this threshold, we employ Boost's Gauss-Kronrod adaptive quadrature integrator. The kinetic spectrum and integrals over $P$ and $z$ also use this integrator. We use a default tolerance of $10^{-12}$ for the kinetic spectrum and $10^{-8}$ for the $P$ and $z$ integrals. We find that the kinetic spectra and gravitational wave spectra converge well given these tolerances, although slightly lower `safe' tolerances are included in \code{config.hpp} for a faster evaluation of the spectrum.

While the sound shell model treats the calculation of the gravitational wave spectrum more carefully than traditional fitting formulas, it tends to be unsuitable for certain transitions. In particular, for very strong phase transitions, fluid velocities become highly relativistic; consequently, non-linearities in the fluid become important to the dynamics of the sound shells as bubbles collide \cite{Cutting:2019zws} and during extreme supercooling \cite{Jinno:2019jhi}. In sufficiently strongly supercooled transitions, or when the coupling to the scalar field is weak, the friction acting on the wall may be insufficient to prevent the wall from reaching a steady state, leading to nearly run-away bubble walls ($\xi_w \to 1$) \cite{Moore:1995si, Huber:2011aa, Bodeker:2009qy, Bodeker:2017cim}. The sound shell model is also known to overestimate the gravitational wave spectrum for deflagrations, where numerical simulations \cite{Cutting:2019zws} show an additional suppression of the kinetic energy of sound waves. Across the parameter space of our scalar singlet extension of the Standard Model, where a first-order phase transition can occur, the strength parameter is $\alpha_N \lesssim 0.1$, and we only consider regions where the wall velocity is subluminal. Therefore, the sound shell model remains a reasonable approximation for this work.
        
\section{Comparison between equation of state models}
\label{sec:Comparison}

To assess the validity of our methods using the exact equation of state, relative to the simplified (bag \& $\mu\nu$) approximations, we first construct fluid profiles and gravitational wave spectra for the benchmark points listed in Table \ref{tbl:BPs}.
We then compare results obtained with all three equations of state in sections \ref{sec:fp-comparison} and \ref{sec:gw-comparison}, respectively. 
Subsequently, we extend this comparison to the full xSM parameter space in section \ref{sec:param-scan}, scanning all regions that admit a first-order phase transition to identify where the simplified equations of state deviate from the exact result in terms of peak amplitude and spectral shape. Finally, in section \ref{sec:snr}, we compute the signal-to-noise ratio (SNR) expected by LISA using the exact equation of state throughout the parameter space.
% 
% table of BPs used
\begin{table}
    \centering
    \footnotesize% fontsize
    \begin{tabular}{|c|c|c|c|c|c|c|c|c|c|c|}
    \hline
    Mode & $\lambda_{hs}$ & $m_s$ [GeV] & $T_*$ [GeV] & $\beta$ [s$^{-1}$] & $\beta/H_*$ & $\xi_w$ & $\alpha_N^{\text{bag}}$ & $\alpha_N^{\mu\nu}$ & $c_{s,+}$ & $c_{s,-}$ \\ \hline
    Def. & 0.826 & 88.75 & 39.78 & $9.52\times10^{-12}$ & 4077.5 & 0.497 & 0.074 & 0.072 & 0.568 & 0.588 \\ \hline
    Hyb. & 0.881 & 106.49 & 62.63 & $4.38\times10^{-12}$ & 761.0 & 0.659 & 0.041 & 0.042 & 0.572 & 0.567 \\ \hline
    Det. & 0.950 & 118.88 & 64.34 & $2.39\times10^{-12}$ & 393.5 & 0.719 & 0.046 & 0.047 & 0.573 & 0.567 \\ \hline
    \end{tabular}
    \caption{Summary of the benchmark points presented in Figures \ref{fig:fluid_profiles} and \ref{fig:gw_spectra} for a $\mathbb{Z}_2$-symmetric extension of the Standard Model. The self-coupling of the scalar field is taken to be $\lambda_s=1$. The phase transition strength parameters $\alpha_N^{\text{bag}}$ and $\alpha_N^{\mu\nu}$ are calculated using \eqref{eq:alpha-bag} and \eqref{eq:alpha-munu}, respectively, and evaluated at the nucleation temperature $T_N \approx T_*$.}\label{tbl:BPs}
\end{table}
\subsection{Fluid profiles}
\label{sec:fp-comparison}
\begin{figure}[t!]
    \centering
    \includegraphics[width=\linewidth]{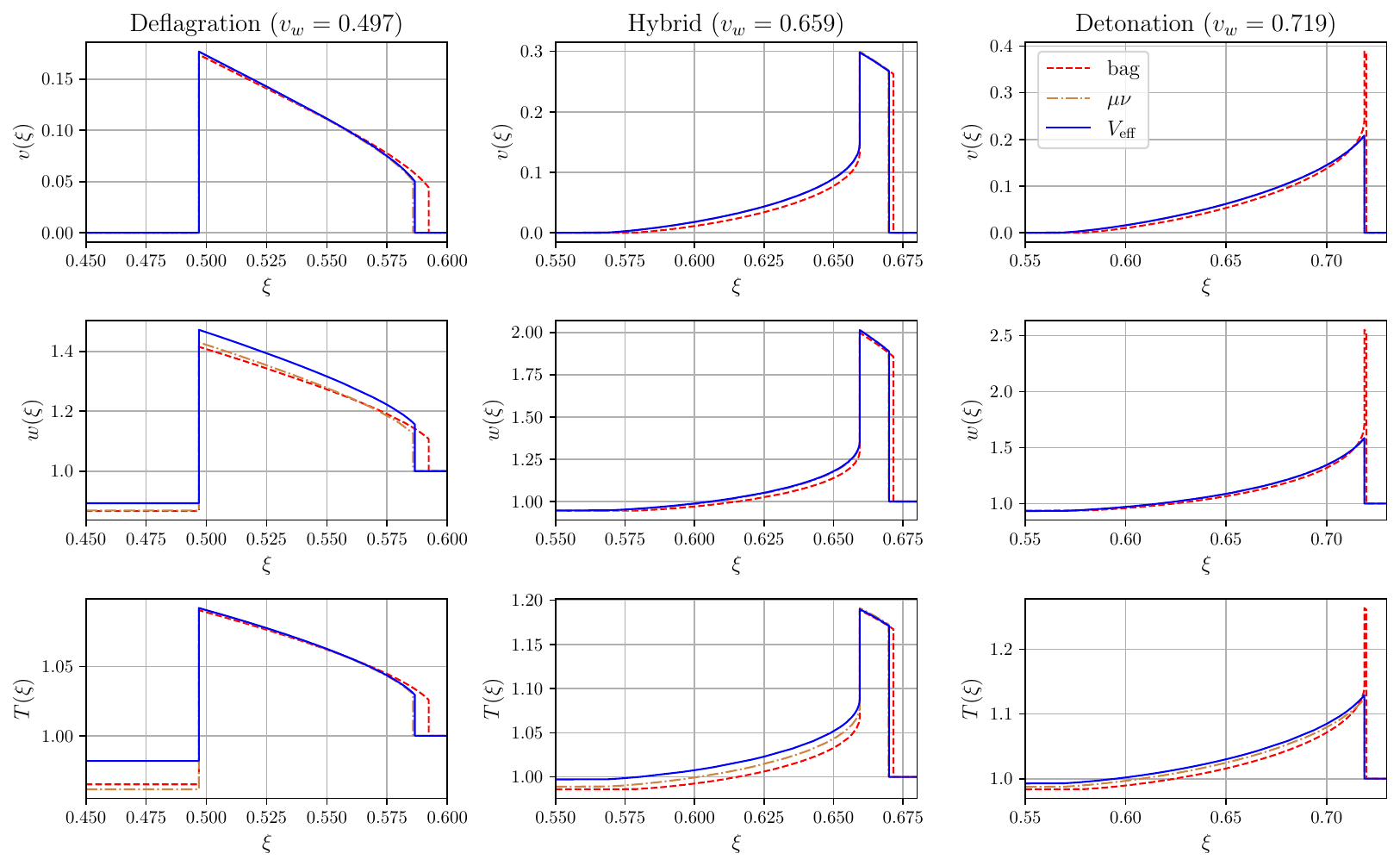}
    \caption{Velocity, enthalpy and temperature profiles of the relativistic plasma for the three types of direct phase transitions. The phase transition parameters for these benchmark points are given in Table \ref{tbl:BPs}. In the detonation profile, the bag equation of state predicts a different hydrodynamic mode (a hybrid, with $|\xi_{\text{sh}} - \xi_w| \ll 1$) to the $\mu\nu$ and exact equations of state.}
    \label{fig:fluid_profiles}
\end{figure}

We construct the fluid profiles during a phase transition described by the effective potential \eqref{eq:Veff-xSM} by solving the fluid equations of motion \eqref{eq:eom-2} and imposing the matching conditions \eqref{eq:matching-conds-2} at the bubble wall and at any shock fronts that appear. We do this using (i) the bag equation of state \eqref{eq:bag-eos}, (ii) the $\mu\nu$ equation of state \eqref{eq:improved-bag-eos}, and (iii) the exact equation of state \eqref{eq:EoS-exact} as determined from the effective potential, as shown in \Cref{fig:fluid_profiles}. The equation of state and phase transition parameters for the effective potential are determined using \code{PhaseTracer} \cite{Athron:2024xrh}, and the wall velocity is calculated using \code{WallGo} \cite{Ekstedt:2024fyq}. 

For the given benchmark points, the $\mu\nu$ model generally approximates the fluid profile from the exact equation of state more accurately than the bag model for deflagration, detonation and hybrid solutions. This is to be expected, given the additional degrees of freedom to specify the sound speed in each vacuum phase. Most notable is the difference in the hydrodynamic mode predicted by each equation of state in the right-most column of \Cref{fig:fluid_profiles}. The bag model predicts a hybrid solution with an extremely narrow profile ($|\xi_{\text{sh}}-\xi_w| \ll 1$), whereas the $\mu\nu$ and exact equations of state predict a detonation solution. This suggests that an accurate determination of the wall velocity and speed of sound is crucial in identifying the correct hydrodynamic description of the fluid. 

Our calculations indicate that the reheating of the fluid inside the bubble (that is within the broken vacuum phase, where $\xi < \xi_w$) is where the largest departure from the radiation-dominated assumption of the bag model ($c_s^2=1/3$) occurs, and is most apparent in the temperature profiles. For the deflagration solution, the position of the shock determined using the bag model differs considerably from the $\mu\nu$ and the exact equations of state. Whilst the $\mu\nu$ model shows an improvement towards the exact equation of state calculation for each solution, it fails to encapsulate the full temperature dependence of the speed of sound within each vacuum phase.

\subsection{Gravitational wave spectra}
\label{sec:gw-comparison}
\begin{figure}[t!]
    \centering
    \begin{tikzpicture}
      \node[anchor=south west, inner sep=0] (img) at (0,0)
        {\includegraphics[width=\linewidth]{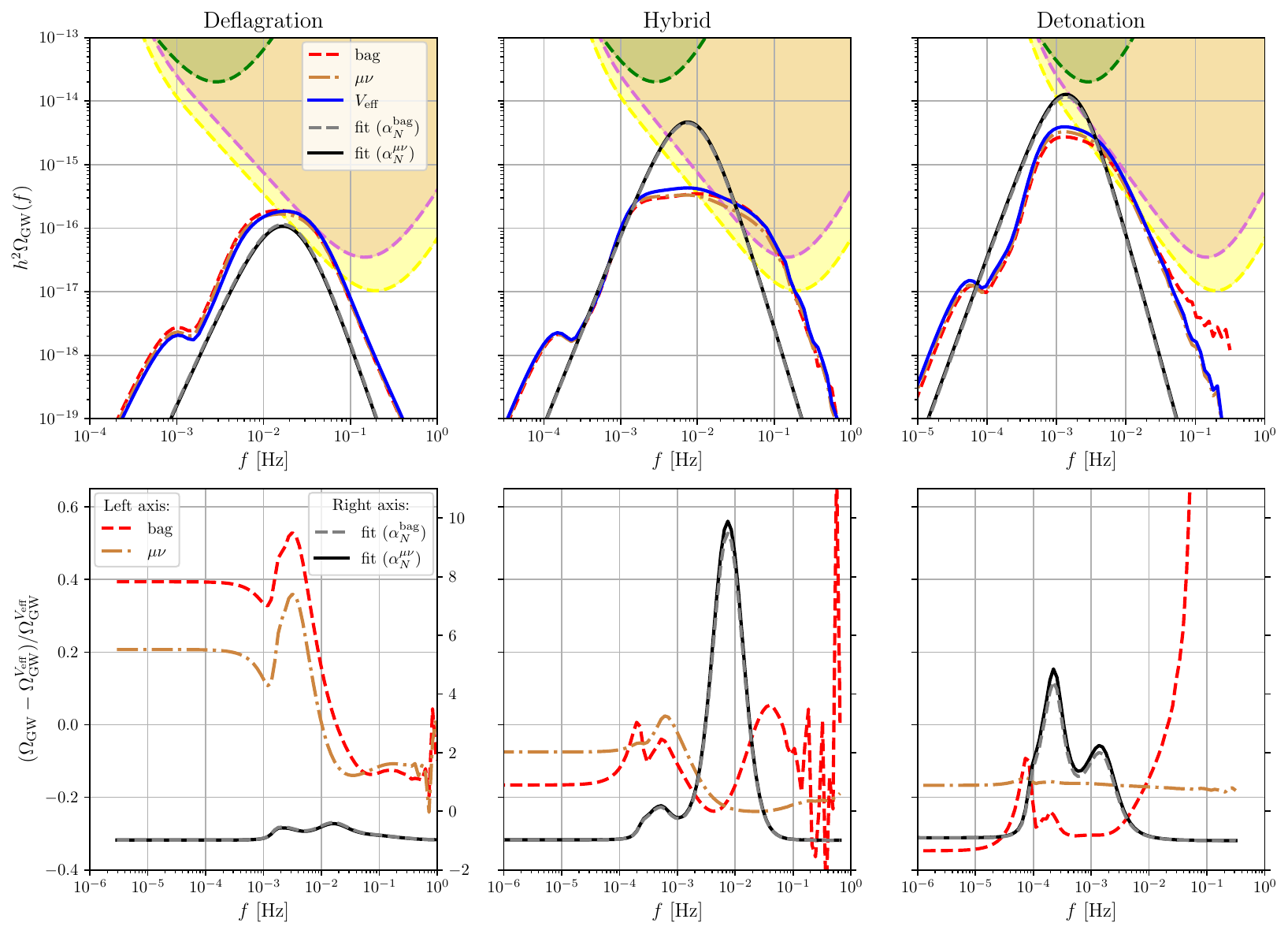}};
      \begin{scope}[x={(img.south east)}, y={(img.north west)}]
        \node at (0.10,0.93) {\footnotesize LISA};
        \draw[->] (0.13,0.93) -- (0.16,0.94);
        \node at (0.12,0.85) {\footnotesize DECIGO};
        \draw[->] (0.17,0.85) -- (0.20,0.85);
        \node at (0.14,0.80) {\footnotesize BBO};
        \draw[->] (0.17,0.80) -- (0.198,0.80);
      \end{scope}
    \end{tikzpicture}
    \caption{In the top row, we show the gravitational wave power spectra due to sound waves, determined using: i) a single broken power law fitting formula \cite{hindmarsh_numerical_2015}, with $\alpha_N$ as determined from the bag and $\mu\nu$ models; and ii) the sound shell model with simplified (bag, $\mu\nu$) and exact ($V_{\text{eff}}$) equations of state. The peak integrated sensitivities for LISA and the proposed future gravitational wave detectors, DECIGO and BBO, are taken from \cite{Schmitz:2020syl} and are shown in the shaded regions. In the bottom row, we show the difference between the gravitational wave spectra, relative to that of the exact equation of state. The bag and $\mu\nu$ spectra are scaled using the left-hand axis, whilst the fitting formula spectra are scaled using the right-hand axis. The phase transition parameters for these benchmark points are given in Table \ref{tbl:BPs}.}
    \label{fig:gw_spectra}
\end{figure}
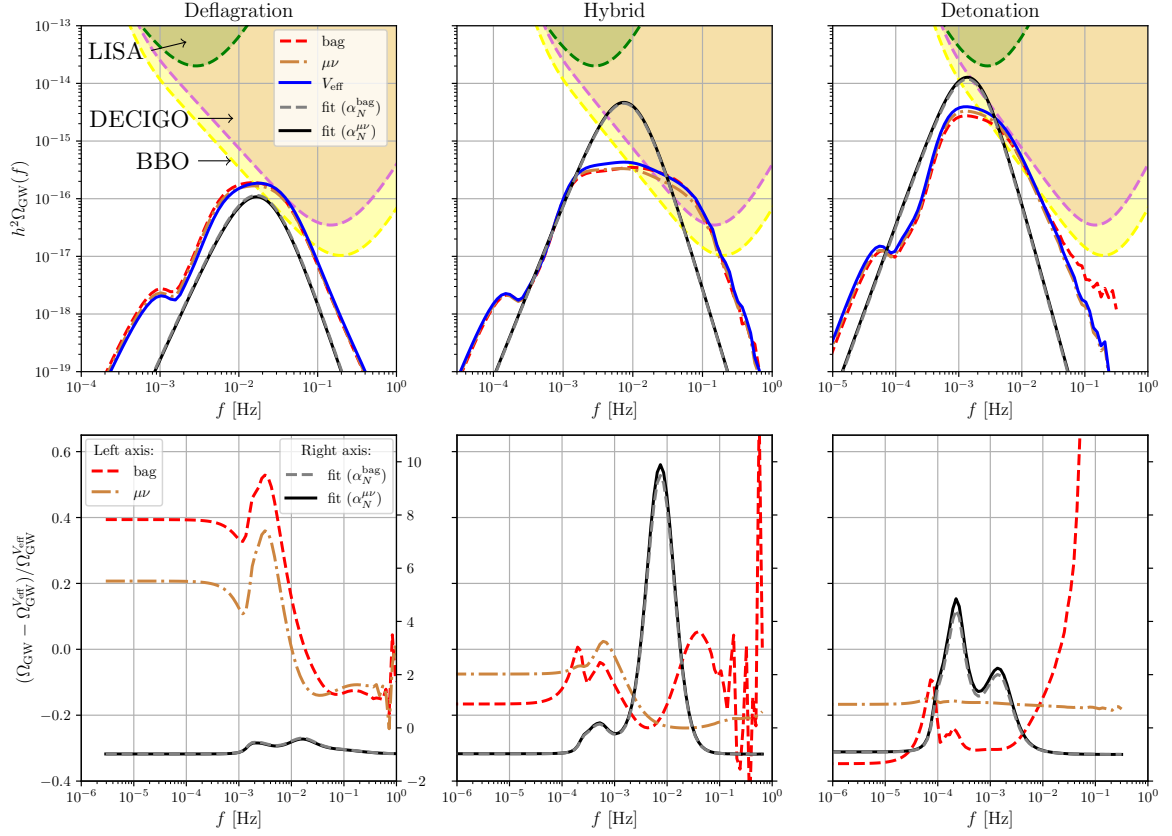

The gravitational wave spectra for the benchmark points in Table \ref{tbl:BPs} were constructed using the sound shell model, as described in section \ref{sec:ssm}, for each equation of state, and are shown in \Cref{fig:gw_spectra}. We take the duration of sound wave production to be the time to develop non-linearities in the fluid\footnote{For an overview of other possible timescales used to estimate the sound wave duration, see, for instance, \cite{Caprini:2019egz}.}, $\delta \tau \approx \delta \tau_{\mathrm{nl}} \sim R_*/\sqrt{\Omega_K}$. Here $\Omega_K$ denotes the average kinetic energy of the fluid, given by \cite{pol_characterization_2024}
\begin{equation}
    \Omega_K \approx \frac{1}{R_*} \int_0^{\infty} \mathrm{d}K \, E_{\mathrm{kin}}(K),
\end{equation}
where $K=k R_*$. For each benchmark point, we find $(\delta \tau/R_*)_\mathrm{def} \sim 12$, $(\delta \tau/R_*)_\mathrm{hyb} \sim 17$, and $(\delta \tau/R_*)_\mathrm{det} \sim 20$.

We first compare our sound shell model spectra to the single broken power law fitting formula \eqref{eq:sw-fit}. For two of the three benchmark points considered, the fitting formula performs poorly in estimating the peak amplitude and, in general, produces a narrower spectrum than the sound shell. In contrast, the double-peak structure predicted by the sound shell model encodes information about both the characteristic length scale, $R_*$, and the thickness of the sound shell, which is determined by the width of the fluid profiles. For deflagration and hybrid solutions with $|\xi_{\text{sh}} - \xi_w| \ll 1$, the sound shell becomes extremely narrow, causing the dominant peak to broaden. This flattening of the peak is a distinctive feature of the sound shell model and is not captured by the fitting formula. Furthermore, the spectra obtained using $\alpha_N^{\mathrm{bag}}$ and $\alpha_N^{\mu\nu}$ in \Cref{fig:gw_spectra} are nearly identical when the fitting formula is employed, whereas the sound shell model predicts a clear difference between the spectra corresponding to the two simplified equations of state. 

The amplitude of the power spectrum is slightly overestimated for the simplified equation of state models in the low-frequency part of the deflagration spectrum, and underestimated for the hybrid and detonation. Furthermore, the peak amplitudes for the hybrid and detonation solutions using the exact equation of state are slightly larger than those from the simplified equation of state calculations. For these benchmark points, we find that the gravitational wave spectra determined using the $\mu\nu$ model of the hydrodynamics tends to agree better with the exact equation of state than the bag model does, as one would expect.

\subsection{xSM parameter scan}
\label{sec:param-scan}
Here, we compute the fluid profiles and gravitational wave spectra across the entire xSM parameter space where a first-order phase transition can occur to examine the validity of the simplified equation of state models. A total of 1037 points (508 deflagrations, 199 hybrids, and 330 detonations) were sampled. Outside of the sampled region, a first-order transition is not possible or does not complete. As before, we take the quartic self-coupling of the scalar field, $\lambda_s$, to be unity, and we vary the scalar field mass, $m_s$\footnote{In practice, we actually vary $\mu_s$, since $m_s$ and $\lambda_{hs}$ are not independent.}, and its coupling to the Higgs, $\lambda_{hs}$. 

A comparison between the hydrodynamic mode predicted by each equation of state model serves as a rudimentary indicator of the overall validity of the simplified equations of state. All deflagrations, as determined using the exact equation of state, agreed between each model. Across the sample, $18\%$ of hybrids were predicted to be deflagrations by the bag model, and $5\%$ of detonations were predicted to be hybrids. Overall, the bag model failed to identify the correct hydrodynamic mode for $\sim 5\%$ of points. The $\mu\nu$ model showed exceptional consistency in determining the hydrodynamic mode, in comparison to the exact equation of state, with only two points ($\lesssim 0.02\%$) of the total sample predicting the incorrect mode.

The hydrodynamic mode entirely depends on the bubble wall velocity, $\xi_w$, and the Jouguet detonation velocity, $v_J^{\text{det}}$, which serves as the boundary between hybrid and detonation solutions for supersonic walls ($\xi_w > c_{s,+}$). The majority of points where the bag model incorrectly predicts the hydrodynamic mode correspond to those where the wall velocity is very close to $v_J^{\text{det}}$. Uncertainties in the determination of $\xi_w$ may therefore also result in the incorrect hydrodynamic mode being assigned to a particular benchmark point for any equation of state model. We therefore warn the reader that precise methods for determining the wall velocity are exceptionally important when modelling the hydrodynamics of phase transitions. In this work, we used the \code{WallGo} code \cite{Ekstedt:2024fyq}, which relies on the bag equation of state, to determine the wall velocity for each sampled point in the xSM parameter space. With this in mind, we aim to incorporate a more general calculation of the wall velocity, using the exact equation of state, in a future release of the \code{HydroGrav} code. 

\begin{figure}[t!]
    \centering
    \includegraphics[width=\textwidth]{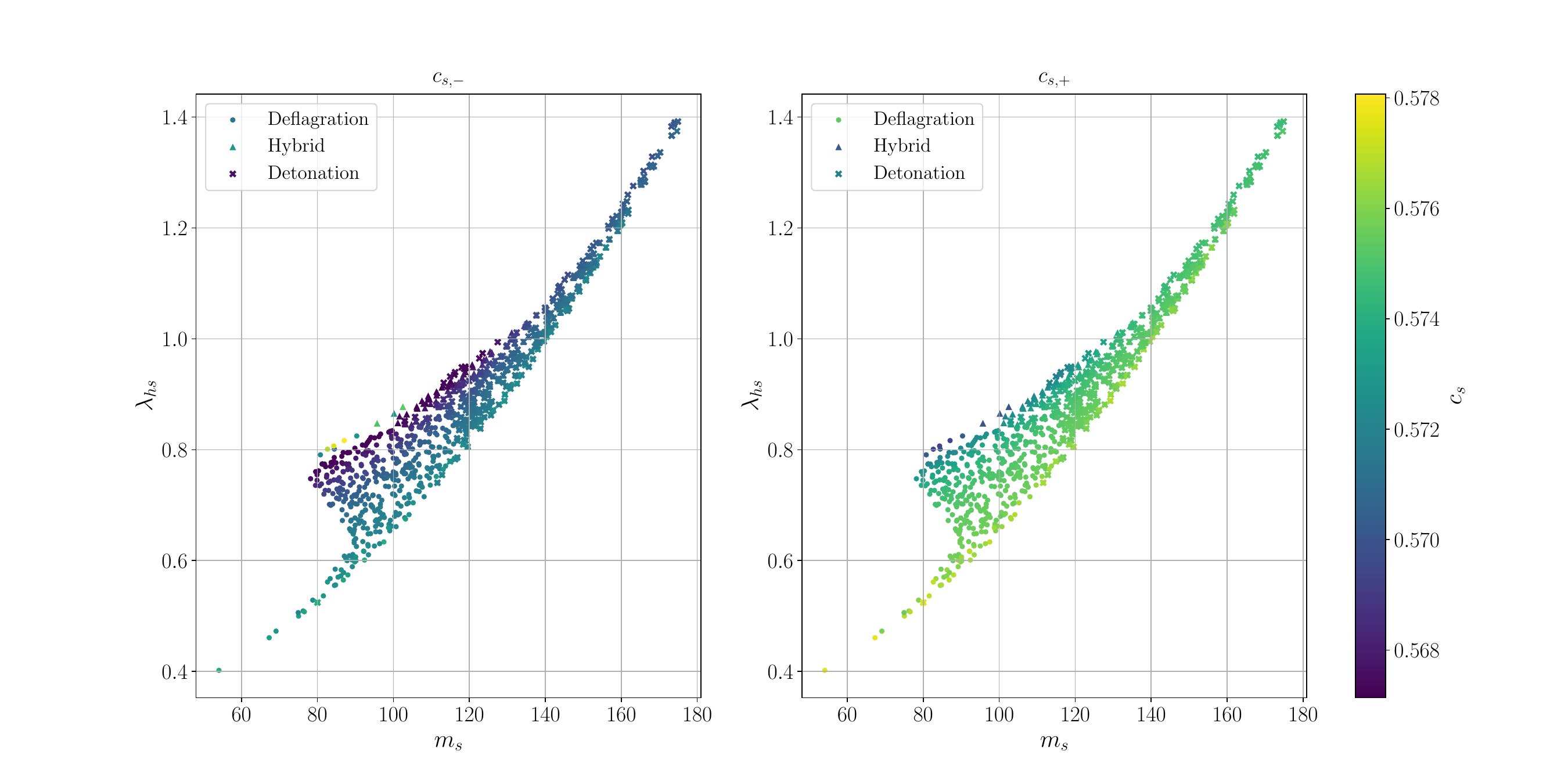}
    \caption{Speed of sound just behind ($-$, broken phase) and just in front ($+$, symmetric phase) of the bubble wall, evaluated at the nucleation temperature. The comparison spans across the xSM parameter space where a first-order phase transition can occur for both deflagration (including hybrid) and detonation solutions.}
    \label{fig:xsm_cs_scan}
\end{figure}

In \Cref{fig:xsm_cs_scan}, we calculate the sound speed in both vacuum phases used to determine the hydrodynamic mode for both the simplified and exact equations of state, estimated by $c_{s,+} = c_{s,s}(T_N)$ and $c_{s,-} = c_{s,b}(T_N)$, where $T_N \approx T_*$. We find a clear gradient in the speed of sound across the parameter space. 
In particular, the speed of sound in the broken phase, $c_{s,-}$, tends to show more variation than in the symmetric phase, $c_{s,-}$. Previously, it was shown \cite{Leitao:2014pda} that the speed of sound in the broken phase departs from the radiation dominated assumption, $c_{s,-}=1/\sqrt{3} \approx 0.577$, in many extensions of the Standard Model that produce a first-order phase transition. Our sound speed analysis demonstrates that this assumption also breaks down, to a lesser degree, in the symmetric phase for some regions of the parameter space, which can be attributed to the reheating of the fluid in front of the wall for deflagration and hybrid solutions. This further reinforces that a departure from the bag model is necessary to construct accurate fluid profiles in some regions of the parameter space. 

\begin{figure}[t!]
    \centering
    \includegraphics[width=\textwidth]{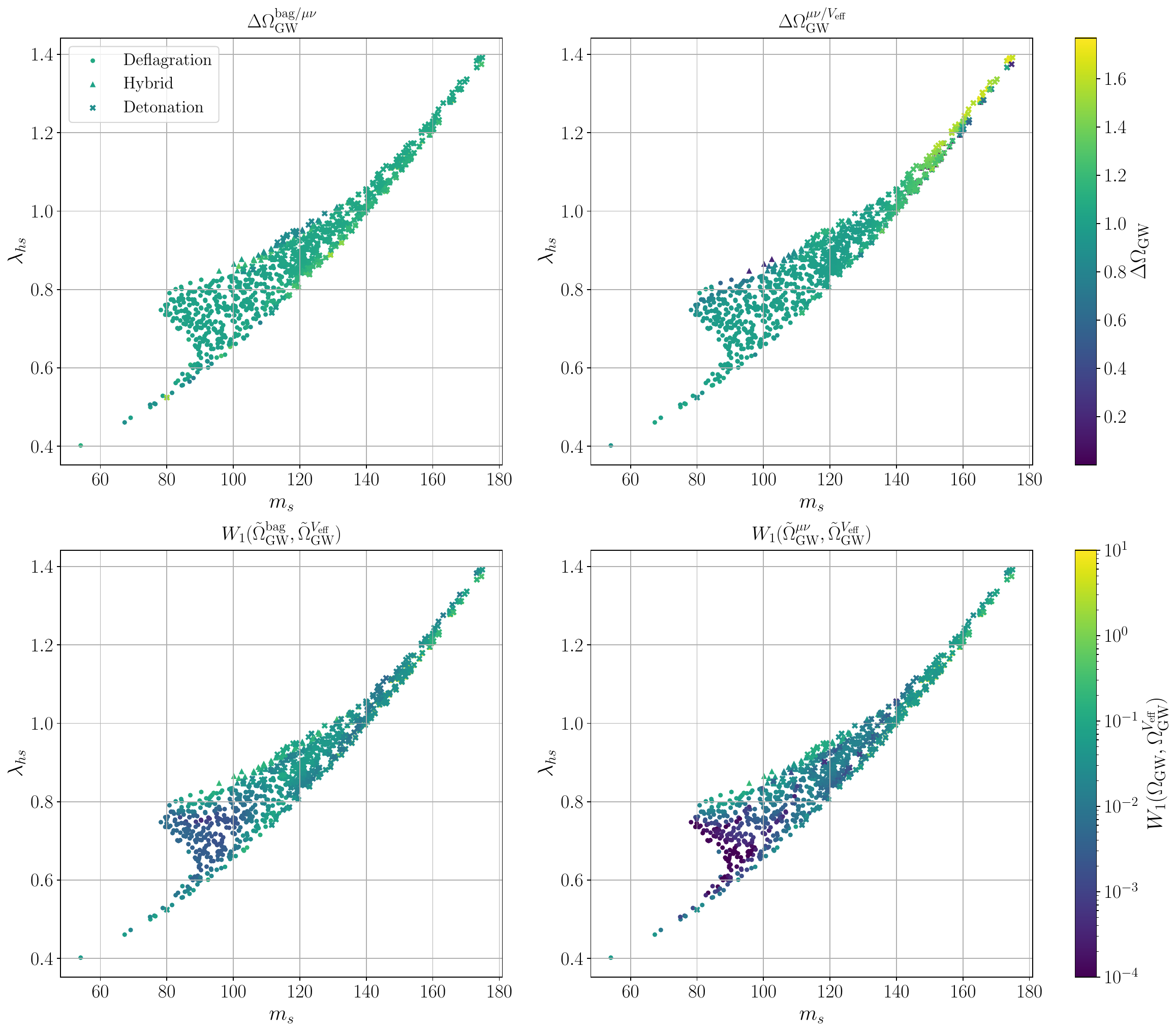}
    \caption{\textit{Top row:} Relative peak amplitude between the gravitational wave spectra determined using the simplified and exact equations of state. The bag model is compared to the $\mu\nu$ model (or improved bag model) in the left panel, and the $\mu\nu$ model is compared to the exact equation of state in the right panel. \textit{Bottom row:} Difference in gravitational wave spectral shape between the simplified and exact equations of state. The Wasserstein distance between the bag ($\mu\nu$) and exact equation of state spectra is shown on the left (right) panel. The comparison spans across the xSM parameter space where a first-order phase transition can occur for both deflagration (including hybrid) and detonation solutions.}
    \label{fig:xsm_gw_scan}
\end{figure}

To perform a more quantitative analysis comparing the gravitational wave spectra between the simplified and exact equations of state, we first calculate the ratio of the peak amplitudes,
\begin{equation}
    \Delta \Omega_{\text{GW}}^{\text{bag}/\mu\nu} = \frac{\Omega_{\text{GW}}^{\text{bag}} (f^{\text{bag}}_{\text{pk}})}{\Omega_{\text{GW}}^{\mu\nu}(f^{\mu\nu}_{\text{pk}})}; \quad \Delta \Omega_{\text{GW}}^{\mu\nu / V_{\text{eff}}} = \frac{\Omega_{\text{GW}}^{\mu\nu} (f^{\mu\nu}_{\text{pk}})}{\Omega_{\text{GW}}^{V_{\text{eff}}}(f^{V_{\text{eff}}}_{\text{pk}})},
\end{equation}
where $f^{\text{eos}}_{\text{pk}}$ is the peak frequency of the gravitational wave spectrum $\Omega_{\text{GW}}^{\text{eos}}$ for each equation of state, and is shown in the top row of \Cref{fig:xsm_gw_scan}. Second, we compute the Wasserstein distance to obtain a scalar measure of the difference in the `shape' of two gravitational wave spectra in the bottom row of \Cref{fig:xsm_gw_scan}. Given two normalised distributions, $\Omega_1$ and $\Omega_2$, the Wasserstein distance is defined by
\begin{equation}
    W_1(\Omega_1, \Omega_2) = \int_{-\infty}^{\infty} \mathrm{d}f \, |F_1(f) - F_2(f)|, \quad F_i(f) = \int_{-\infty}^f \mathrm{d}f' \, \Omega_i(f'), \label{eq:wasserstein-distance}
\end{equation}
where $F_i(f)$ is the cumulative distribution function of $\Omega_i$. In particular, we are interested in the difference between the gravitational wave spectra determined using either of the simplified equations of state and the exact equation of state, so we compute $W_1(\tilde{\Omega}_{\text{GW}}^{\text{bag, }\mu\nu}, \tilde{\Omega}_{\text{GW}}^{V_{\text{eff}}})$, where the tilde ($\sim$) denotes the normalised spectrum with $\tilde{\Omega}^{\text{eos}}_{\text{GW}}(f^{\text{eos}}_{\text{pk}})=1$. A statistical summary for the parameter scan is shown in \Cref{tab:stat-sum}. 

\begin{table}[t!]
\centering
\footnotesize
\begin{tabular}{|l|lllll|}
\hline
Metric & Mode & Mean & Median & $\sigma$ & $N$ \\ \hline
\multirow{4}{*}{$\Delta\Omega_{\mathrm{GW}}^{\mathrm{bag}/\mu\nu}$} & Deflagration & 1.0158 & 1.0162 & 0.0617 & 508 \\
 & Hybrid & 1.1453 & 1.1170 & 0.1746 & 199 \\
 & Detonation & 1.0657 & 1.0489 & 0.0809 & 330 \\
 & Global & 1.0565 & 1.0328 & 0.1101 & 1037 \\ \hline
\multirow{4}{*}{$\Delta\Omega_{\mathrm{GW}}^{\mu\nu/V_{\mathrm{eff}}}$} & Deflagration & 0.9437 & 0.9873 & 0.1518 & 508 \\
 & Hybrid & 0.6603 & 0.9610 & 0.4295 & 199 \\
 & Detonation & 1.1107 & 1.0388 & 0.2185 & 330 \\
 & Global & 0.9425 & 0.9892 & 0.2933 & 1037 \\ \hline
\multirow{4}{*}{$W_1(\tilde{\Omega}_{\mathrm{GW}}^{\mathrm{bag}}, \tilde{\Omega}_{\mathrm{GW}}^{V_{\mathrm{eff}}})$} & Deflagration & 0.1242 & 0.0150 & 0.4368 & 508 \\
 & Hybrid & 1.0802 & 0.1958 & 1.2876 & 199 \\
 & Detonation & 0.0565 & 0.0327 & 0.1961 & 330 \\
 & Global & 0.2861 & 0.0340 & 0.7571 & 1037 \\ \hline
\multirow{4}{*}{$W_1(\tilde{\Omega}_{\mathrm{GW}}^{\mu\nu}, \tilde{\Omega}_{\mathrm{GW}}^{V_{\mathrm{eff}}})$} & Deflagration & 0.1148 & 0.0058 & 0.4500 & 508 \\
 & Hybrid & 1.0162 & 0.1101 & 1.2546 & 199 \\
 & Detonation & 0.0476 & 0.0184 & 0.1747 & 330 \\
 & Global & 0.2664 & 0.0167 & 0.7376 & 1037 \\ \hline
$c_{s,-}$ & Global & 0.5706 & 0.5710 & 0.0015 & 1037 \\ \hline
$c_{s,+}$ & Global & 0.5749 & 0.5751 & 0.0010 & 1037 \\ \hline
\end{tabular}
\caption{Statistical summary for the ratio of peak amplitudes and difference in spectral shape across the sampled region of the xSM parameter space. The $\mu\nu$ approximation reproduces the peak amplitude to within $1\%$ of $\Omega_{\text{GW}}^{V_{\text{eff}}}(f^{V_{\text{eff}}}_{\text{pk}})$ for $36/15/11$\% (def./hyb./det.) of the sampled points, whereas the bag model does so for $36/4/5$\%.}
\label{tab:stat-sum}
\end{table}

As one would expect, the bottom row of \Cref{fig:xsm_gw_scan} shows that the $\mu\nu$ model replicates the shape of the gravitational wave spectra from the exact equation of state slightly better than the bag model, with a smaller Wasserstein distance across the entire parameter space. This is especially true in the region where $80 \, \mathrm{GeV} \lesssim m_s \lesssim 100 \, \mathrm{GeV}$. The variation in the spectral shape between different equation of state models further reinforces that basic single broken power law fitting formulas are not suitable to accurately describe the sound wave contribution to the gravitational wave spectrum of a first-order phase transition. 

The bag and $\mu\nu$ models exhibit nearly identical peak amplitudes across the parameter space, with a much smaller standard deviation than that obtained when comparing the $\mu\nu$ and the exact equations of state. In both comparisons, hybrid solutions are found to exhibit the largest differences in peak amplitude. This trend is further reflected in the Wasserstein distance for hybrid solutions, whose median value is at least an order of magnitude larger than those for deflagrations and detonations.

The large difference between the median and mean Wasserstein distance for both deflagrations and hybrids suggests that points where the exact equation of state spectrum differs considerably from the simplified equations of state have driven up the mean. For hybrid solutions, with $m_s \gtrsim 140$ GeV, and deflagration solutions, with $130 \, \mathrm{GeV} \lesssim m_s \lesssim 150 \, \mathrm{GeV}$, the spectra computed from the exact equation of state have much larger peak amplitudes than the $\mu\nu$ model, and $W_1 \sim 10$ for both bag and $\mu\nu$ models. This likely occurs due to numerical artefacts in constructing the fluid profiles using the exact equation of state for solutions where the fluid profile is extremely narrow ($|\xi_{\text{sh}} - \xi_w| \ll 1$) and the matching conditions at the shock cannot be accurately determined. 

The regions where hybrid solutions are found generally coincide with darker regions in \Cref{fig:xsm_cs_scan}, where we observe the largest deviations in the sound speed from $c_{s}^{\mathrm{bag}}=1/\sqrt{3} \approx 0.577$ in both vacuum phases. In contrast, detonations are found primarily for scalar masses $m_s>100$ GeV, where the sound speed is very close to $c_s^{\mathrm{bag}}$, and the difference in spectral shape between the three equation of state models is much smaller than that of the other hydrodynamic modes.

\subsection{Signal-to-noise ratio}
\label{sec:snr}
\begin{figure}[t!]
    \centering
    \includegraphics[width=0.8\linewidth]{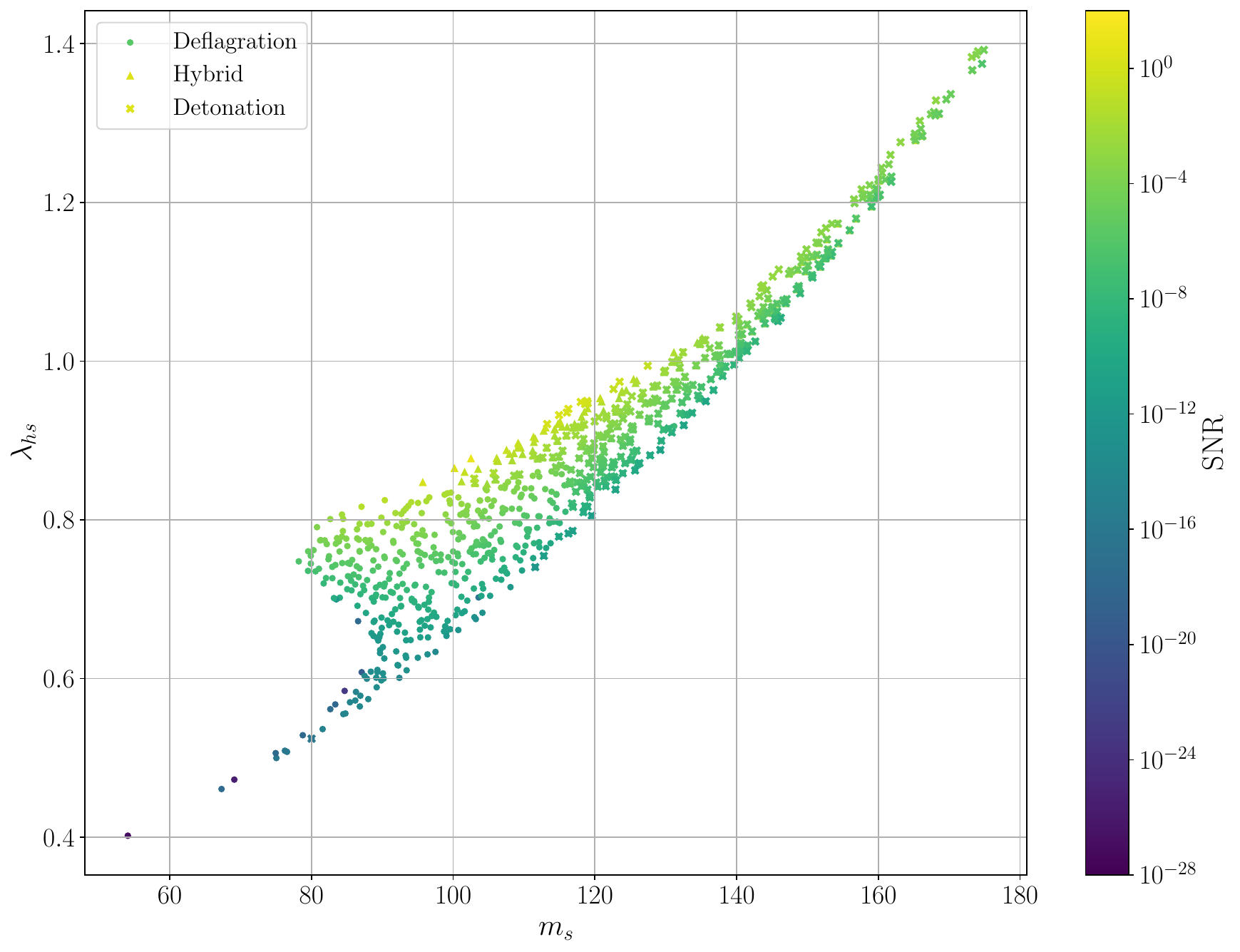}
    \caption{Signal-to-noise ratio (SNR) of the gravitational wave spectrum, as measured by LISA across its 4-year lifetime (approx. 3 years of data collection), using the exact equation of state across the xSM parameter space where a first-order phase transition can occur.}
    \label{fig:xsm_gw_scan_snr}
\end{figure}
To identify the regions of parameter space that are most likely to produce detectable gravitational wave signals, we compute the signal-to-noise ratio (SNR)~\cite{Thrane:2013oya} associated with the spectrum, $h^2\Omega_{\rm GW}(f)$, using
\begin{equation}
    \text{SNR} = \sqrt{n_\text{det} t_\text{obs} \int_{f_\text{min}}^{f_\text{max}} \mathrm{d}f \, \left[ \frac{h^2 \Omega_\text{GW} (f)}{h^2 \Omega_\text{noise} (f)} \right]^2}.
\end{equation}
Here, $n_\text{det}$ is the effective number of detectors, which distinguishes between auto-correlation ($n_\text{det}=1$) and cross-correlation ($n_\text{det}=2$) measurements. The setup of LISA will allow for an auto-correlation measurement \cite{Cutler:1997ta}, whereas the future proposed detectors DECIGO and BBO will effectively form two-detector networks capable of cross-correlation measurements \cite{Schmitz:2020syl}. The duration of the observation run is $t_\text{obs}$, given in years. LISA is expected to operate with a duty cycle of $\mathcal{D}=0.75$, which implies that its planned four-year mission will yield an effective observing time of approximately three years of gravitational wave data~\cite{LISA_SciRD}. 
The detector bandwidth spans the frequency range $[f_\text{min}, f_\text{max}]$. For LISA, the detector is designed to be sensitive to the frequency band between $0.1$ mHz and $1$ Hz \cite{LISA:2024hlh}. Finally, $h^2 \Omega_\text{noise} (f)$ is the noise power spectrum of the detector. For a detailed overview describing the calculation of the noise spectrum, we refer the reader to \cite{Schmitz:2020syl}. Gravitational wave signals whose SNR is greater than the threshold $\text{SNR}_\text{thr}$ are detectable, with the detection threshold for LISA expected to be $\text{SNR}_\text{thr}=10$ \cite{Caprini:2015zlo}. 

The SNR, as measured by LISA over a four-year mission and determined using the exact equations of state, is shown in \Cref{fig:xsm_gw_scan_snr}. Across the parameter space of our model, the SNR is found to lie below the detection threshold of LISA, with a maximum value of approximately unity. However, we emphasise that the purpose of our work is not to examine the discovery potential; rather, it is to demonstrate a fully model-dependent pipeline.
Additionally, the accumulation of theoretical uncertainties associated with the calculation of a first order phase transition, ranging from the construction of the effective potential to the prediction of the gravitational wave spectrum, may substantially alter the overall scale of the signal-to-noise ratio across the parameter space. A comprehensive study of the theoretical uncertainties affecting gravitational wave predictions from first-order phase transitions is deferred to future work.

% Conclusions
\section{Conclusions}
\label{sec:concl}
We presented \code{HydroGrav}, a \code{C++} framework for computing self-similar fluid profiles and gravitational wave spectra from cosmological first-order phase transitions using either simplified equations of state or the exact equation of state obtained directly from a finite-temperature effective potential.  The code combines model-dependent relativistic hydrodynamics with the sound shell model, enabling a fully consistent connection between the microscopic thermodynamics of an underlying particle-physics model, the resulting fluid motion, and the predicted gravitational wave signal.  In addition to supporting the widely used bag and $\mu\nu$ equations of state, \code{HydroGrav} provides a general framework for studying the impact of equation-of-state effects on gravitational wave predictions.

Using a $\mathbb{Z}_2$-symmetric real singlet extension of the Standard Model as a case study, we compared fluid profiles and gravitational wave spectra obtained using the bag, $\mu\nu$, and exact equations of state.  We find that, for this model, the $\mu\nu$ equation of state generally reproduces the exact hydrodynamic solutions more accurately than the bag equation of state.  The largest deviations from the radiation-dominated approximation occur in the broken phase, where the temperature dependence of the thermodynamic quantities leads to departures from the constant sound speed assumption underlying the bag equation of state.  In some regions of parameter space, these differences are sufficiently large to alter the predicted hydrodynamic mode, demonstrating that simplified equations of state can occasionally lead to qualitatively different descriptions of the plasma dynamics.

Extending the comparison across the full parameter space of the model, we find that the gravitational wave spectra predicted by the bag or $\mu\nu$ and the exact equation of state may differ significantly in both the spectral shape and peak amplitude.  For the singlet extension considered here, these results suggest that the $\mu\nu$ approximation captures the dominant thermodynamic effects relevant for gravitational wave production. However, this conclusion should not be regarded as universal.  

The degree to which simplified equations of state remain reliable is expected to depend on the thermodynamic structure of the underlying model, and larger deviations may arise in theories with more complicated thermal histories or stronger departures from radiation domination.  For the $\mathbb{Z}_2$-symmetric real singlet extension, across $76/50/47$\% (def./hyb./det.) of the parameter space, the peak amplitude predicted by the $\mu\nu$ model agrees with the exact result to within 5\%.

Finally, we evaluated the signal-to-noise ratio expected for LISA across the parameter space of the model using the exact equation of state.  For the parameter regions considered in this work, the predicted signals remain below the nominal detection threshold of LISA, although theoretical uncertainties associated with the effective potential, bubble-wall dynamics, hydrodynamics, and gravitational wave production remain important and deserve further investigation.

The framework developed here provides a foundation for systematic studies of these effects.  
As gravitational wave observatories move towards precision measurements of cosmological signals, fully model-dependent treatments of phase-transition hydrodynamics will become increasingly important for establishing robust connections between observations and the underlying particle physics.

\acknowledgments
F.L. and W.S. are supported by the Commonwealth through an Australian Government Research Training Program Scholarship, \href{https://doi.org/10.82133/C42F-K220}{doi.org/10.82133/C42F-K220}.
X.W. and C.B. are supported by Australian Research Council grants DP220100643 and LE250100010.

\appendix

\section{Examples}
\label{app:examples}
\code{HydroGrav} can generate fluid profiles and gravitational wave spectra for the multiple equations of state listed in the work above and for any set of thermal parameters. In addition, it provides functionality to evaluate the velocity (or kinetic) power spectrum. Below, we provide a short code example to generate a gravitational wave spectrum. We indicate how to modify this for the bag, $\mu \nu$, and generic equations of state. More detailed examples can be found in the \code{examples} directory of the repository.
\begin{lstlisting}[style=cppstyle]
#include "hydrograv.hpp"

int main() 
{
    const double T0 = 2.34914e-13; // 2.725 K
    const double Ts = 52.9772;
    const double H0 = 1.44328e-42; // 67.8 km/s/Mpc
    const double Hs = 4.34679e-15;
    const double g0 = 3.91;
    const double gs = 106.75;
    
    const PhaseTransition::Universe un(T0, Ts, g0, gs, H0, Hs);

    const double vw = 0.8;
    const double alpha = 0.1;
    const double beta = 1e-12;
    const double Rs = std::pow(8*M_PI, 1./3.) * vw / beta;
    const std::string nuc_type = "exp";
    
    const PhaseTransition::PTParams_Bag params_bag(vw, alpha, Ts, beta, Rs, nuc_type, un);

    const std::vector<double> kRs_vals = logspace(-3.0, 3.0, 100);
    const double dtau = 10.0 * Rs;
    
    Spectrum::PowerSpec OmegaGW = Spectrum::GWSpec(kRs_vals, params_bag, dtau);

    const std::vector<double> frequency_values = OmegaGW.freq();
    const std::vector<double> momentum_values = OmegaGW.K();
    const std::vector<double> amplitude_values = OmegaGW.P();
    
    return;
}
\end{lstlisting}
The example begins by initialising the \code{Universe} and \code{PTParams} objects needed to generate the spectra. The former contains the Hubble rate, reference temperature, and relativistic degrees of freedom for the transition. A `\code{0}' is included to indicate present-day values, whilst an `\code{s}' (starred) indicates the values taken at the transition temperature. All dimensionful quantities are in units of GeV. Likewise, we must initialise an instance of the \code{PTParams} object. In the example above, we use the default constructor for the bag model, which requires $\xi_w$, $\alpha$, $T_\star$, $\beta$, $R_*$, the nucleation type, and the \code{Universe} object. The nucleation type can be either exponential (\code{"exp"}) or simultaneous (\code{"sim"}). The initialisation above defaults to the bag model equation of state. For users wishing to instead use the $\mu \nu$ model, \code{PTParams\_Bag} should further be initialised with the sound speed:
\begin{lstlisting}[style=cppstyle]
    const double csq_plus = 1/3;
    const double csq_minus = 1/3 - 0.01;
    
    const PhaseTransition::PTParams_Bag params_bag(vw, alpha, Ts, beta, Rs, nuc_type, un, csq_plus, csq_minus);
\end{lstlisting}
To forego the bag equation of state entirely and instead use the fully generalised equation of state, the \code{PTParams} object must be initialised with an instance of the \code{EquationOfState} object. This contains vectors of the temperature and the pressure and energy density in both the symmetric and broken phases, and is initialised as such:
\begin{lstlisting}[style=cppstyle]
    const std::vector<double> T;
    const std::vector<double> ps;
    const std::vector<double> pb;
    const std::vector<double> es;
    const std::vector<double> eb;
    
    PhaseTransition::EquationOfState eos(T, ps, pb, es, eb)
\end{lstlisting}
The generic \code{PTParams\_Veff} object can then be initialised using:
\begin{lstlisting}[style=cppstyle]
    const PhaseTransition::PTParams_Veff params_veff(vw, alpha, Ts, beta, Rs, nuc_type, un, eos);
\end{lstlisting}
In addition to the phase transition parameters, the \code{GWSpec} object requires a set of $K=k R_*$ values for which to generate the spectrum. We create this in the code above using the \code{logspace} helper function, analogous to its \code{numpy} counterpart. Furthermore, the sound wave duration, $\delta \tau$, should also be passed into \code{GWSpec}. This describes the duration of the gravitational wave source, and is typically given in terms of the mean bubble separation, $R_*$. As done in this work, $\delta \tau$ can be estimated by the time to develop non-linearities in the fluid, $\delta \tau_{\mathrm{nl}}$, using:
\begin{lstlisting}[style=cppstyle]
    const Hydrodynamics::FluidProfile profile(params_veff); // or params_bag
    const double dtau = get_nl_timescale(profile);
\end{lstlisting}
Once the \code{GWSpec} object has been initialised, the user can access vectors of the calculated frequency, $K$ values, and amplitude by accessing the \code{freq}, \code{K}, and \code{P} member attributes, respectively.

% Bibliography
\bibliographystyle{JHEP}
\bibliography{biblio-1.bib}

\end{document}